\documentclass[a4paper,12pt]{article}
\usepackage{epsfig}
\usepackage{color}
\usepackage[numbers,sort&compress]{natbib}
\usepackage{multirow}

\newlength{\dinwidth}
\newlength{\dinmargin}
\newcommand{\spur}[1]{\not\! #1 \,}

\begin{document}
\title{Probing the R-parity violating
supersymmetric effects in the exclusive $c \to d/s\ell\nu_\ell$  decays}
\author{ Ru-Min Wang\thanks{E-mail:ruminwang@gmail.com},~Jie Zhu,~Jin-Huan Sheng,~Mo-Lin Liu,~Yuan-Guo Xu
  \\
{\scriptsize {\it College of Physics and Electronic
Engineering, Xinyang Normal University,
 Xinyang, Henan 464000, China
}}
 }
 \maketitle
\vspace{-0.4cm}

\begin{abstract}
A lot of branching ratios of the exclusive $c \to d/s\ell\nu_\ell$ ($\ell=e,\mu$) decays have been quite accurately measured by CLEO-c, BELLE, BABAR, BES(I,II,III), ALEPH and MARKIII collaborations.
We probe the R-parity violating supersymmetric effects  in the exclusive  $c \to
 d/s\ell\nu_\ell$ decays. From  the latest experimental measurements, we obtain
new upper limits on the relevant R-parity violating coupling parameters within the decays, and many upper limits are obtained for the first time.
Using the constrained new parameter spaces, we predict the R-parity violating effects on the observables,
 which have not been measured or have not been well measured yet.
We find that the  R-parity violating effects due to slepton exchange could be large
on the branching ratios of $D_{d/s}\to e\nu_e$ decays and the normalized forward-backward asymmetries of $D_{u/d}\to \pi/K \ell\nu_\ell$ as well as  $D_s\to K \ell\nu_\ell$ decays, and all branching ratios of the relevant semileptonic $D$ decays are sensitive to  squark exchange couplings.
 Our results in this work could be used to probe new physics effects in
the leptonic decays as well as the semileptonic decays, and will correlate with searches for
direct supersymmetric signals at LHC and BESIII.
\end{abstract}

\newpage

\section{Introduction}

The $ c\to d/s \ell\nu_\ell$ transitions have played a central role for the most precise measurements of CKM matrix  elements $V_{cd}$ and $V_{cs}$ for a long time. These rare charmed decays also have received a lot of  attention, since they are very promising for investigating the standard model (SM) and searching for new physics (NP) beyond it.
The 26 charmed decays,  $D_d\to \ell\nu_\ell$, $D_u\to \pi\ell\nu_\ell$, $D_d\to\pi\ell\nu_\ell$, $D_s\to K\ell\nu_\ell$, $D_u\to \rho\ell\nu_\ell$, $D_d\to\rho \ell\nu_\ell$, $D_s\to K^* \ell\nu_\ell$, $D_s\to \ell\nu_\ell$, $D_u\to K\ell\nu_\ell$, $D_d\to K \ell\nu_\ell$, $D_u\to K^{*}\ell\nu_\ell$, $D_d\to K^{*} \ell\nu_\ell$ and $D_s\to \phi \ell\nu_\ell$, are dominated by $c \to d/s\ell\nu_\ell$ transitions.
Many  collaborations, such as
 BSEIII \cite{Ma:2014dha,Li:2012tr,Huang:2012qc,Ablikim:2013uvu}, CLEO-c \cite{Besson:2009uv,Huang:2005iv,Coan:2005iu,Yelton:2009aa,Eisenstein:2008aa,Briere:2010zc,Alexander:2009ux,Ecklund:2009aa} and BELLE \cite{Widhalm:2006wz,Widhalm:2007ws}, BABAR \cite{delAmoSanchez:2010jg,Aubert:2008rs}, BESII \cite{Ablikim:2006bw},
BES \cite{Ablikim:2004ry,Ablikim:2006ah,Ablikim:2004ej,Ablikim:2004ku},  ALEPH \cite{Heister:2002fp} and MARK-III \cite{Adler:1989rw}, have studied the exclusive $c \to d/s\ell\nu_\ell$ decays, and a lot of
branching ratios have been quite accurately measured by them.
Present experimental measurements  are in good agreement with the SM predictions, and they give us an opportunity to  disprove NP or find bounds over NP models beyond the SM.

The  exclusive $c \to d/s\ell\nu_\ell$ decays have been studied extensively in the SM
and its various extensions (see for instance Refs. \cite{Barranco:2014bva,Barranco:2013tba,Akeroyd:2009tn,Dobrescu:2008er,Akeroyd:2007eh,Fajfer:2006uy,Fajfer:2005ug,Fajfer:2004mv,Akeroyd:2003jb,Akeroyd:2002pi}).
In the present study, we will analyze these  decays in supersymmetry  (SUSY) without R-parity.
We will obtain eight new upper limits on the relevant supersynmmetric coupling parameters that satisfy all of the experimental data from the relevant charmed decays.
Using the constrained new parameter spaces, we will predict
the R-parity violating (RPV) effects on the branching ratios, the differential branching ratios
and the normalized forward-backward asymmetries of charged leptons.
 Our results imply that the constrained  RPV couplings due to slepton exchange have great effects on   $\mathcal{B}(D_{d/s}\to e\nu_e)$, and they  could obviously enhance the allowed ranges of $\mathcal{A}_{FB}(D \to P \ell\nu_\ell)$.   Nevertheless, the RPV contributions due to squark exchange couplings could enhance the predictions of all semileptonic branching ratios, which are very sensitive to the relevent RPV coupling products.

This paper is schemed as follows: In section 2, we introduce the theoretical frame
of the exclusive $c \to d/s\ell\nu_\ell$
decays in SUSY without $R$-parity. In section 3,  we deal with the numerical results. We display the constrained parameter spaces which satisfy
all the available experimental data, and then we use the constrained parameter spaces to predict the RPV effects on other quantities,
which have not been measured or have not been well measured yet. Section 4 contains our summary and conclusion.

\section{The exclusive $c \to d/s\ell\nu_\ell$  decays in SUSY without R-parity}
In the SM,  the $c \to d/s\ell\nu_\ell$ processes are mediated by  a virtual $W$ boson exchange, and the relevant four fermion effective Hamiltonian is
\begin{eqnarray}
\mathcal{H}^{SM}_{eff}(\bar{c} \to \bar{d}_k \ell^+_m\nu_{\ell_n})=
\frac{G_F}{\sqrt{2}}V^*_{cd_k}(\bar{d}_k\gamma_{\mu}(1-\gamma_5)c)(\bar{\nu}_{\ell_n}\gamma^{\mu}(1-\gamma_5)\ell_m).\label{SMHamiltonian}
\end{eqnarray}

In the most general superpotential
of SUSY, the RPV superpotential
is given by \cite{Weinberg:1981wj}
\begin{eqnarray}
\mathcal{W}_{\spur{R_p}}&=&\mu_i\hat{L}_i\hat{H}_u+\frac{1}{2}
\lambda_{[ij]k}\hat{L}_i\hat{L}_j\hat{E}^c_k+
\lambda'_{ijk}\hat{L}_i\hat{Q}_j\hat{D}^c_k+\frac{1}{2}
\lambda''_{i[jk]}\hat{U}^c_i\hat{D}^c_j\hat{D}^c_k, \label{RPVsuperpotential}
\end{eqnarray}
where $\hat{L}$ and $\hat{Q}$ are the SU(2)-doublet lepton and quark
superfields, $\hat{E}^c$, $\hat{U}^c$ and $\hat{D}^c$ are the
singlet superfields, while $i$, $j$ and $k$ are generation indices
and $c$ denotes a charge conjugate field.
From Eq. (\ref{RPVsuperpotential}), one can get the relevant R-parity breaking
part of the Lagrangian of  $\bar{c} \to \bar{d}_j \ell^+_m\nu_{\ell_n}$  \cite{Petrov:2007gp,Chemtob:2004xr}
\begin{eqnarray}
\mathcal{L}_{eff}^{\spur{R_p}}(\bar{c} \to \bar{d}_k \ell^+_m\nu_{\ell_n})&=&-\sum_{i,j,k}\tilde{\lambda}'_{ijk}\left[\tilde{\ell}_{iL}\bar{d}_{kR}u_{jL}+\tilde{d}^*_{kR}\bar{\ell}^c_{iR}u_{jL}\right]\nonumber\\
&&+\sum_{i,j,k}\lambda'_{ijk}\left[\tilde{d}^*_{kR}\bar{\nu}^c_{\ell_iR}d_{jL}\right]+\sum_{i,j,k}\lambda_{ijk}\left[\tilde{\ell}_{jL}\bar{\ell}_{kR}\nu_{\ell_iL}\right],\label{RPVL}
\end{eqnarray}
with $\tilde{\lambda}_{irk}\equiv\sum_nV^*_{rn}\lambda_{ink}$, and $V_{rn}$ is the SM CKM matrix element. Noted that (s)down-down-(s)neutrino vertices have the weak eigenbasis couplings
$\lambda'$, while charged (s)lepton-(s)down-(s)up vertices have the up quark mass eigenbasis
couplings $\tilde{\lambda}'$. Very often in the literature (see e.g. \cite{Golowich:2006gq,Chen:2007dg,Nandi:2006qe,Bhattacharyya:1998be,Kundu:2004cv}), one neglects the difference between
$\lambda'$ and $\tilde{\lambda}'$, based on the fact that diagonal elements of the CKM matrix dominate over nondiagonal
ones.

\begin{figure}[t]
\begin{center}
\includegraphics[scale=1]{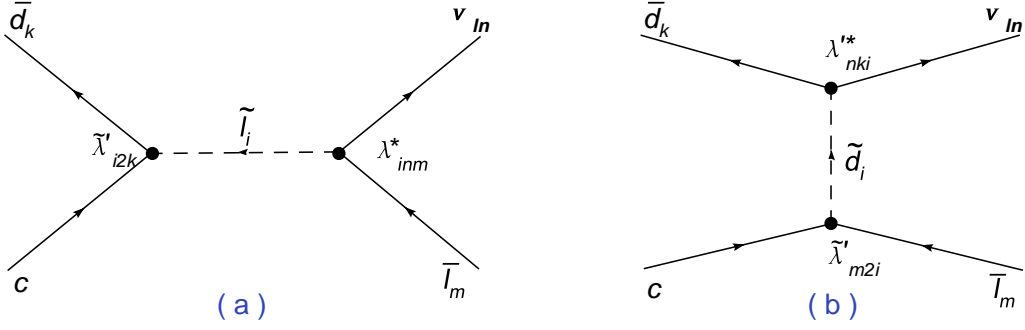}
\end{center}
\vspace{-0.4cm}
 \caption{The RPV contributions to the exclusive  $\bar{c} \to \bar{d}_k \ell^+_m\nu_{\ell_n}$
decays due to slepton and squark exchange.}
 \label{fig:feynman}
\end{figure}

 In terms of Eq.(\ref{RPVL}), we can obtain the relevant four fermion effective Hamiltonian
for the $\bar{c} \to \bar{d}_j \ell^+_m\nu_{\ell_n}$ processes with RPV
couplings due to the squark and slepton exchange
\begin{eqnarray}
\mathcal{H}_{eff}^{\spur{R_p}}(\bar{c} \to \bar{d}_k \ell^+_m\nu_{\ell_n})&=&
-\sum_i\frac{\lambda'^*_{nki}\tilde{\lambda}'_{m2i}}{8m^2_{\tilde{d}_{iR}}}(\bar{d}_k\gamma_{\mu}(1-\gamma_5)c)(\bar{\nu}_{\ell_n}\gamma^{\mu}(1-\gamma_5)\ell_m)\nonumber\\
&&+\sum_i\frac{\lambda^*_{inm}\tilde{\lambda}'_{i2k}}{4m^2_{\tilde{ \ell}_{iL}}}(\bar{d}_k(1-\gamma_5)c)(\bar{\nu}_{\ell_n}(1+\gamma_5)\ell_m).
\end{eqnarray}
 And the corresponding RPV feynman diagrams for the  $\bar{c} \to \bar{d}_k \ell^+_m\nu_{\ell_n}$ processes are displayed in Fig. \ref{fig:feynman}.

 Then we can obtain the total effective Hamiltonian for the $\bar{c} \to \bar{d}_k \ell^+_m\nu_{\ell_n}$ in the RPV SUSY
\begin{eqnarray}
\mathcal{H}_{eff}(\bar{c} \to \bar{d}_k \ell^+_m\nu_{\ell_n})&=&
\mathcal{H}_{eff}^{SM}(\bar{c} \to \bar{d}_k \ell^+_m\nu_{\ell_n})+\mathcal{H}_{eff}^{\spur{R_p}}(\bar{c} \to \bar{d}_k \ell^+_m\nu_{\ell_n}). \label{total effective Hamiltonian}
\end{eqnarray}
Based on the effective Hamiltonian in Eq. (\ref{total effective Hamiltonian}), we will give the expressions of physical quantities for the RPV SUSY later in detail. In the following expressions and numerical analysis, we will keep the masses of the charged leptons, but ignore all neutrino masses.

\subsection{$D_{d/s}\to \ell\nu_\ell$ decays }
Purely leptonic decays are the simplest and the cleanest decay modes of the pseudoscalar
charged $D^+$ meson, and the decay amplitude of $D^+_{d_k}\to \ell^+\nu_\ell$ can be obtained in the terms of Eq.(\ref{total effective Hamiltonian})
\begin{eqnarray}
\mathcal{M}^{\spur{R_p}}(D^+_{d_k}\to \ell^+_m\nu_{\ell_n})&=&
<\ell^+_m\nu_{\ell_n} | \mathcal{H}_{eff}(\bar{c} \to \bar{d}_k \ell^+_m\nu_{\ell_n})|D^+_{d_k} > \nonumber\\
&=&\left[\frac{G_F}{\sqrt{2}}V^*_{cd_k}-\sum_i\frac{\lambda'^*_{nki}\tilde{\lambda}'_{m2i}}{8m^2_{\tilde{d}_{iR}}}\right]
<0|\bar{d}_k\gamma_{\mu}(1-\gamma_5)c|D^+_{d_k}>(\bar{\nu}_{\ell_n}\gamma^{\mu}(1-\gamma_5)\ell_m)\nonumber\\
&&+\sum_i\frac{\lambda^*_{inm}\tilde{\lambda}'_{i2k}}{4m^2_{\tilde{\ell}_{iL}}}<0|\bar{d}_k(1-\gamma_5)c|D^+_{d_k}>(\bar{\nu}_{\ell_n}(1+\gamma_5)\ell_m).
\end{eqnarray}
After using the definitions of $D$ meson decay constant \cite{Grinstein:1992qt}
\begin{eqnarray}
<0|\bar{d}_k\gamma_{\mu} \gamma_5 c|D^+(p)>=if_{D}p_\mu,
\end{eqnarray}
and
\begin{eqnarray}
 <0|\bar{d}_k \gamma_5 c|D^+(p)>=-if_D\mu_{D_{q}}~~\mbox{with}~~\mu_{D_{d_k}}\equiv \frac{m^2_{D_{d_k}}}{\bar{m}_c+\bar{m}_{d_k}},
\end{eqnarray}
we get the branching ratio for $D^+_{d/s}\to \ell^+\nu_\ell$
\begin{eqnarray}
\mathcal{B}^{\spur{R_p}}(D^+_{d_k}\to \ell^+_m\nu_{\ell_n})&=&
\left|\frac{G_F}{\sqrt{2}}V^*_{cd_k}-\sum_i\frac{\lambda'^*_{nki}\tilde{\lambda}'_{m2i}}{8m^2_{\tilde{d}_{iR}}}
+\sum_i\frac{\lambda^*_{inm}\tilde{\lambda}'_{i2k}}{4m^2_{\tilde{\ell}_{iL}}}\frac{\mu_{D_{d_k}}}{m_{\ell_m}}\right|^2\nonumber\\
&&\times\frac{\tau_{D_{d_k}}}{4\pi}f^2_{D_{d_k}}m_{D_{d_k}}m^2_{\ell_m}\left[ 1-\frac{m^2_{\ell_m}}{m^2_{D_{d_k}}}\right].
\end{eqnarray}

\subsection{$D \to P \ell\nu_\ell$  $(P=\pi,K)$ decays }

 In the terms of Eq.(\ref{total effective Hamiltonian}), $D \to P \ell^+\nu_\ell$ decay amplitude can be written as
\begin{eqnarray}
\mathcal{M}^{\spur{R_p}}(D \to P\ell^+_m\nu_{\ell_n})&=&
\left<P\ell^+_m\nu_{\ell_n}|\mathcal{H}_{eff}(\bar{c}\to\bar{d}_k\ell^+_m\nu_{\ell_n})|D\right> \nonumber\\
&=&\left[\frac{G_F}{\sqrt{2}}V^*_{cd_k}-\sum_i\frac{\lambda'^*_{nki}\tilde{\lambda}'_{m2i}}{8m^2_{\tilde{d}_{iR}}}\right]
\left<P|\bar{d}_k\gamma_{\mu}(1-\gamma_5)c|D\right>(\bar{\nu}_{\ell_n}\gamma^{\mu}(1-\gamma_5)\ell_m)\nonumber\\
&&+\sum_i\frac{\lambda^*_{inm}\tilde{\lambda}'_{i2k}}{4m^2_{\tilde{\ell}_{iL}}}\left<P|\bar{d}_k(1-\gamma_5)c|D\right>(\bar{\nu}_{l_n}(1+\gamma_5)\bar{\ell}_m).
\end{eqnarray}
Using the $D \to P$ transition form factors \cite{Wu:2006rd}
\begin{eqnarray}
c_P\left<P(p)\left|\bar{d}_k\gamma_{\mu}c\right|D(p_D)\right>&=&f^P_+(s)(p+p_D)_{\mu}+\left[f^P_0(s)-f^P_+(s)\right]\frac{m^2_D-m^2_P}{s}q_{\mu},\\
c_P\left<P(p)\left|\bar{d}_kc\right|D(p_D)\right>&=&f^P_0(s)\frac{m^2_D-m^2_P}{\bar{m}_c-\bar{m}_{d_k}},
\end{eqnarray}
with the factor $c_P$ accounts for the flavor content of particles ($c_P=\sqrt{2}$ for $\pi^0$, and $c_P=1$ for $\pi^-, K^0, K^-$ ) and $s=q^2$ ($q=p_D-p$), the differential branching ratio for $ D \to P \ell^+_m\nu_{\ell_n}$ is
\begin{eqnarray}
\frac{d\mathcal{B}^{\spur{R_p}}(D \to P \ell^+_m \nu_{\ell_n})}{ds dcos\theta}=\frac{\tau_D\sqrt{\lambda_P}}{2^7\pi^3m^3_Dc^2_P}\left(1-\frac{m^2_{\ell_m}}{s}\right)^2\left[N^P_0+N^P_1 cos\theta+N^P_2cos^2\theta\right],
\end{eqnarray}
with
\begin{eqnarray}
N^P_0&=&\left|\frac{G_F}{\sqrt{2}}V^*_{cd_k}-\sum_i\frac{\lambda'^*_{nki}\tilde{\lambda}'_{m2i}}{8m^2_{\tilde{d}_{iR}}}\right|^2[f^P_+(s)]^2\lambda_P
+\left|\frac{G_F}{\sqrt{2}}V^*_{cd_k}-\sum_i\frac{\lambda'^*_{nki}\tilde{\lambda}'_{m2i}}{8m^2_{\tilde{d}_{iR}}}\right.\nonumber\\
&&+\left.
\sum_i\frac{\lambda^*_{inm}\tilde{\lambda}'_{i2k}}{4m^2_{\tilde{\ell}_{iL}}}\frac{s}{m_{\ell_m}(\bar{m}_c-\bar{m}_{d_k})}\right|^2m^2_{\ell_m}[f^P_0(s)]^2\frac{(m^2_D-m^2_P)^2}{s},\\
N^P_1&=&\left\{\left|\frac{G_F}{\sqrt{2}}V^*_{cd_k}-\sum_i\frac{\lambda'^*_{nki}\tilde{\lambda}'_{m2i}}{8m^2_{\tilde{d}_{iR}}}\right|^2
+Re\left[\left(\frac{G_F}{\sqrt{2}}V^*_{cd_k}-\sum_i\frac{\lambda'^*_{nki}\tilde{\lambda}'_{m2i}}{8m^2_{\tilde{d}_{iR}}}\right)^{\dagger}\right.\right.\nonumber\\
&&\left.\left.\times\sum_i\frac{\lambda^*_{inm}\tilde{\lambda}'_{i2k}}{4m^2_{\tilde{\ell}_{iL}}}\frac{s}{m_{\ell_m}(\bar{m}_c-\bar{m}_{d_k})}\right]\right\}
2m^2_{\ell_m}f^P_0(s)f^P_+(s)\sqrt{\lambda_P}\frac{(m^2_D-m^2_P)}{s},\\
N^P_2&=&-\left|\frac{G_F}{\sqrt{2}}V^*_{cd_k}-\sum_i\frac{\lambda'^*_{nki}\tilde{\lambda}'_{m2i}}{8m^2_{\tilde{d}_{iR}}}\right|^2[f^P_+(s)]^2\lambda_P\left(1-\frac{m^2_{\ell_m}}{s}\right),
\end{eqnarray}
where $\theta$ is the angle between the momentum of $D$ meson and the charged lepton in the c.m. system of $\ell-\nu$, and the kinematic factor
$\lambda_P=m^4_D+m^4_P+s^2-2m^2_Dm^2_P-2m^2_Ds-2m^2_Ps$.

Here, we give the definition of the normalized forward-backward (FB) asymmetry of charged lepton, which is more useful from the experimental point of view,
\begin{eqnarray}
\bar{\mathcal{A}}_{FB}=\frac{\int^{+1}_0\frac{d^2\mathcal{B}}{ds dcos\theta}dcos\theta-\int^0_{-1}\frac{d^2\mathcal{B}}{ds dcos\theta}dcos\theta}
{\int^{+1}_0\frac{d^2\mathcal{B}}{ds dcos\theta}dcos\theta+\int^0_{-1}\frac{d^2\mathcal{B}}{ds dcos\theta}dcos\theta}.
\end{eqnarray}
Explicitly, for $ D \to P \ell^+\nu_\ell$ the normalized FB asymmetry is
\begin{eqnarray}
\bar{\mathcal{A}}_{FB}(D\to P\ell^+\nu_\ell)=\frac{N^P_1}{2N^P_0+2/3N^P_2}.
\end{eqnarray}

\subsection{$ D \to V \ell^+\nu_l$  $(V= \rho,K^{*},\phi)$ decays }

From Eq.(\ref{total effective Hamiltonian}), $D\to V\ell^+\nu_\ell$  decay amplitude can be written as
\begin{eqnarray}
\mathcal{M}^{\spur{R_p}}(D\to V\ell^+_m\nu_{\ell_n})&=&
\left<V\ell^+_m \nu_{\ell_n}|\mathcal{H}_{eff}(\bar{c} \to \bar{d}_k \ell^+_m \nu_{\ell_n})|D\right> \nonumber\\
&=&\left[\frac{G_F}{\sqrt{2}}V^*_{cd_k}-\sum_i\frac{\lambda'^*_{nki}\tilde{\lambda}'_{m2i}}{8m^2_{\tilde{d}_{iR}}}\right]
\left<V|\bar{d}_k\gamma_{\mu}(1-\gamma_5)c|D\right>(\bar{\nu}_{\ell_n}\gamma^{\mu}(1-\gamma_5)\ell_m)\nonumber\\
&&+\sum_i\frac{\lambda^*_{inm}\tilde{\lambda}'_{i2k}}{4m^2_{\tilde{\ell}_{iL}}}\left<V|\bar{d}_k(1-\gamma_5)c|D\right>(\bar{\nu}_{\ell_n}(1+\gamma_5)\ell_m).
\end{eqnarray}
In terms of the $D\to V$ form factors \cite{Wu:2006rd}
\begin{eqnarray}
c_V\left<V(p,\varepsilon^*)\left|\bar{d}_k\gamma_{\mu}(1-\gamma_5)c\right|D(p_D)\right>
&=&\frac{2V^V(s)}{m_D+m_V}\epsilon_{\mu\nu\alpha\beta}\varepsilon^{*\nu}p^\alpha_Dp^\beta\nonumber\\
&&-i\left[\varepsilon^*_\mu(m_D+m_V)A^V_1(s)-(p_D+p)_\mu(\varepsilon^*.p_D)\frac{A^V_2(s)}{m_D+m_V}\right]\nonumber\\
&&+iq_\mu(\varepsilon^*.p_D)\frac{2m_V}{s}[A^V_3(s)-A^V_0(s)],\\
c_V\left<V(p,\varepsilon^*)\left|\bar{d}_k\gamma_5c\right|D(p_D)\right>
&=&-i\frac{\varepsilon^*.p_D}{m_D}\frac{2m_Dm_V}{\bar{m}_c+\bar{m}_{d_k}}A^V_0(s),
\end{eqnarray}
where $c_V=\sqrt{2}$ for $\rho^0$, $c_V=1$ for $\rho^- ,K^{*0}, K^{*-},\phi$ and with the relation
 $A^V_3(s)=\frac{m_D+m_V}{2m_V}A^V_1(s)-\frac{m_D-m_V}{2m_V}A^V_2(s)$ and $A^V_0(s)=A^V_3(s)-A^{'V}_3(s)$, we have
\begin{eqnarray}
\frac{d\mathcal{B}^{\spur{R_p}}(D \to V \ell^+_m \nu_{\ell_n})}{ds dcos\theta}=\frac{\tau_D\sqrt{\lambda_V}}
{2^7\pi^3m^3_Dc^2_V}\left(1-\frac{m^2_{\ell_m}}{s}\right)^2\left[N^V_0+N^V_1 cos\theta+N^V_2cos^2\theta\right],
\end{eqnarray}
with
\small{
\begin{eqnarray}
N^V_0&=&\left|\frac{G_F}{\sqrt{2}}V^*_{cd_k}-\sum_i\frac{\lambda'^*_{nki}\tilde{\lambda}'_{m2i}}{8m^2_{\tilde{d}_{iR}}}\right|^2
\left\{[A^V_1(s)]^2\left(\frac{\lambda_{V}}{4m^2_V}+(m^2_{\ell_m}+2s)\right)(m_D+m_V)^2\right.\nonumber\\
&&+[A^V_2(s)]^2\frac{\lambda_{V}^2}{4m_V^2(m_D+m_V)^2}+[V^V(s)]^2\frac{\lambda_{V}}{(m_D+m_V)^2}(m_{\ell_m}^2+s)\nonumber\\
&&\left.-A_1^V(s)A_2^V(s)\frac{\lambda_{V}}{2m^2_V}(m^2_D-s-m_V^2)\right\}\nonumber\\
&&+\left|\frac{G_F}{\sqrt{2}}V^*_{cd_k}-\sum_i\frac{\lambda'^*_{nki}\tilde{\lambda}'_{m2i}}{8m^2_{\tilde{d}_{iR}}}+
\sum_i\frac{\lambda^*_{inm}\tilde{\lambda}'_{i2k}}{4m^2_{\tilde{\ell}_{iL}}}\frac{s}{m_{\ell_m}(\bar{m}_c+\bar{m}_{d_k})}\right|^2
[A^v_0(s)]\frac{m^2_{\ell_m}}{s}\lambda_{V},\\
N^V_1&=&\left\{\left|\frac{G_F}{\sqrt{2}}V^*_{cd_k}-\sum_i\frac{\lambda'^*_{nki}\tilde{\lambda}'_{m2i}}{8m^2_{\tilde{d}_{iR}}}\right|^2\right.\nonumber\\
&&\left.+Re\left[\left(\frac{G_F}{\sqrt{2}}V^*_{cd_k}-\sum_i\frac{\lambda'^*_{nki}\tilde{\lambda}'_{m2i}}{8m^2_{\tilde{d}_{iR}}}\right)^{\dagger}
\sum_i\frac{\lambda^*_{inm}\tilde{\lambda}'_{i2k}}{4m^2_{\tilde{\ell}_{iL}}}\frac{s}{m_{\ell_m}(\bar{m}_c-\bar{m}_{d_k})}\right]\right\}\nonumber\\
&&\times\left[A^V_0(s)A^V_1(s)\frac{m_{\ell_m}^2(m_D+m_V)(m^2_D-m^2_V-s)\sqrt{\lambda_{V}}}{sm_V}-A^V_0(s)A^V_2(s)\frac{m^2_{\ell_m}\lambda^{\frac{3}{2}}_{V}}{sm_V(m_D+m_V)}\right]\nonumber\\
&&+\left|\frac{G_F}{\sqrt{2}}V^*_{cd_k}-\sum_i\frac{\lambda'^*_{nki}\tilde{\lambda}'_{m2i}}{8m^2_{\tilde{d}_{iR}}}\right|^2A^V_1(s)V^V(s)4s\sqrt{\lambda_{V}},\\
N^V_2&=&-\left|\frac{G_F}{\sqrt{2}}V^*_{cd_k}-\sum_i\frac{\lambda'^*_{nki}\tilde{\lambda}'_{m2i}}{8m^2_{\tilde{d}_{iR}}}\right|^2\left(1-\frac{m^2_{\ell_m}}{s}\right)
\lambda_{V}\left\{[A^V_1(s)]^2\frac{(m_D+m_V)^2}{4m_V^2}+[V^V(s)]^2\frac{s}{(m_D+m_V)^2}\right.\nonumber\\
&&\left.+[A^V_2(s)]^2\frac{\lambda_{V}}{4m_V^2(m_D+m_V)^2}-A^V_1(s)A^V_2(s)\frac{m_D^2-m_V^2-s}{2m_V^2}\right\},
\end{eqnarray}}
where $\lambda_V=m^4_D+m^4_V+s^2-2m^2_Dm^2_V-2m^2_Ds-2m^2_Vs$.

The normalized FB asymmetry of $D \to V {\ell}^+\nu_\ell$ can be written as
\begin{eqnarray}
\bar{\mathcal{A}}_{FB}(D \to V {\ell}^+\nu_\ell)=\frac{N^V_1}{2N^V_0+2/3N^V_2}.
\end{eqnarray}

\section{Numerical Results and Discussions}
In this section, we summarize our numerical results and analysis of RPV couplings in the  exclusive $\bar{c}\to \bar{d}/\bar{s}\ell^+\nu_\ell$  decays.
When we study the effects due to SUSY without R-parity,
 we consider only one new coupling at one time, neglecting the interferences between
different new couplings, but keeping their interferences with the SM
amplitude. The input parameters are collected in the Appendix.  To be conservative, the input parameters varied
randomly within $1\sigma$  variance and the experimental bounds at
90\% confidence level (CL) will be used to constrain parameters of the relevant new
couplings.

 \subsection{The exclusive $\bar{c}\to \bar{d}e^+\nu_e$ decays}
 There are two RPV coupling products,
$\lambda^*_{i11}\tilde{\lambda}'_{i21}$ due to slepton exchange and $\lambda'^*_{11i}\tilde{\lambda}'_{12i}$ due to squark exchange,  contributing to seven exclusive
$\bar{c}\to \bar{d}e^+\nu_e$ decay modes,
 $D^+_d\to e^+\nu_e$, $D^0_u\to\pi^-e^+\nu_e$,
 $D^+_d\to\pi^0e^+\nu_e$, $D^+_s\to K^0e^+\nu_e$,  $D^0_u\to\rho^-e^+\nu_e$,
 $D^+_d\to\rho^0e^+\nu_e$ and $D^+_s\to K^{*0}e^+\nu_e$.
All  relevant semilpetonic branching ratios of  the exclusive $\bar{c}\to \bar{d}e^+\nu_e$ decays have been accurately measured by BSEIII \cite{Ma:2014dha,Li:2012tr,Huang:2012qc}, CLEO-c \cite{Besson:2009uv,Huang:2005iv,Coan:2005iu,Yelton:2009aa} and Belle \cite{Widhalm:2006wz}, furthermore,  the pureleptonic branching ratios of  $D^+_d\to e^+\nu_e$ is upperlimited by CLEO-c \cite{Eisenstein:2008aa}.
  Their average values from PDG \cite{PDG2014} and  corresponding experimental bound at 90\% CL are given in the second column of Table \ref{tab:cde}. Moreover, the SM prediction values with $1\sigma$ error ranges for
the input parameters  are listed in  the third column of Table \ref{tab:cde}.
 \begin{table}[b]
\caption{Branching ratios of the exclusive $\bar{c}\to \bar{d}e^+\nu_e $ decays are in units of $10^{-3}$ except
  for  branching ratio of $D^+_d\to e^+\nu_{e}$ is in units of $10^{-8}$. ``a" denotes the experimental data and ``b" denotes the corresponding experimental bound at 90\% CL. }
\begin{center}
\begin{tabular}{lcccc}
\hline\hline
 Observable& Exp. data & SM predictions& SUSY w/$\lambda^*_{i11}\tilde{\lambda}'_{i21}$ & SUSY w/$\lambda'^*_{11i}\tilde{\lambda}'_{12i}$ \\\hline
$\mathcal{B}(D^+_d\to e^+\nu_e) $&$<880$&$[0.71,1.02]$&$<880$&$[0.49,1.24]$\\\hline
$\mathcal{B}(D^0_u\to \pi^{-}e^+\nu_e)$&$^{(2.89\pm0.08)^a}_{[2.76,~3.02]^b}$&$[1.86,6.37]$&$[2.89,3.02]$&$[2.90,3.02]$\\\hline
$\mathcal{B}(D^+_d\to \pi^0e^+\nu_e)$&$^{(4.05\pm0.18)^a}_{[3.75,~4.35]^b}$&$[2.39,8.21]$&$[3.75,3.92]$&$[3.75,3.91]$\\\hline
$\mathcal{B}(D^+_s\to K^0e^+\nu_e)$&$^{(3.7\pm1.0)^a}_{[2.06,~5.34]^b}$&$[2.36,5.32]$&$[2.36,5.32]$&$[2.06,5.34]$\\\hline
$\mathcal{B}(D^0_u\to \rho^{-}e^+\nu_e)$&$^{(1.9\pm0.4)^a}_{[1.24,~2.56]^b}$&$[1.49,2.10]$&$[1.49,2.10]$&$[1.24,2.21]$\\\hline
$\mathcal{B}(D^+_d\to \rho^0e^+\nu_e)$&$^{(2.2\pm0.4)^a}_{[1.54,~2.86]^b}$&$[1.93,2.72]$&$[1.93,2.72]$&$[1.60,2.86]$\\\hline
$\mathcal{B}(D^+_s\to K^{*0}e^+\nu_e)$&$^{(1.8\pm0.7)^a}_{[0.65,~2.95]^b}$&$[1.92,2.74]$&$[1.92,2.74]$&$[1.35,2.92]$\\\hline
\hline
\end{tabular}
\end{center}
\label{tab:cde}
\end{table}
The theoretical
predictions of relevant  branching fractions are consistent with experimental data at the present level of precision, and we can constrain the relevant NP parameter spaces by these $D$ decays.

\begin{figure}[t]
\begin{center}
\includegraphics[scale=0.65]{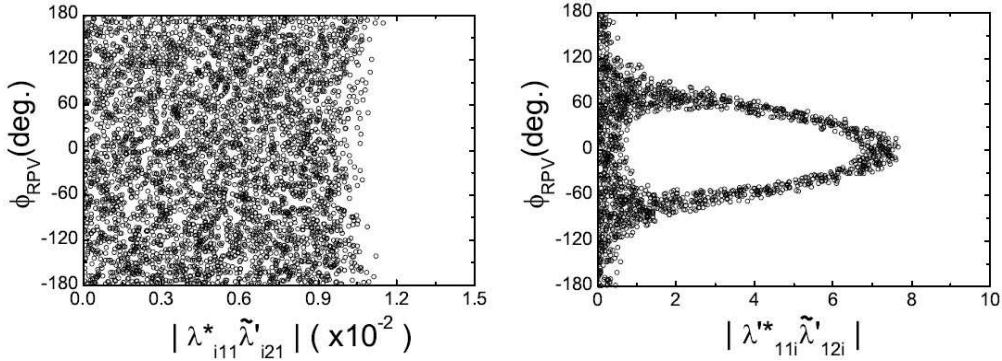}
\end{center}
\vspace{-0.4cm}
 \caption{The allowed RPV parameter spaces from the exclusive $\bar{c}\to \bar{d}e^+\nu_e$ decays  at 90\% CL with  $500$ GeV sfermion mass.}
 \label{fig:boundscde}
\end{figure}

Our bounds for
$\lambda^*_{i11}\tilde{\lambda}'_{i21}$ and $\lambda'^*_{11i}\tilde{\lambda}'_{12i}$  from
the 90\% CL experimental data are demonstrated in
Fig. \ref{fig:boundscde}.
We can see that the moduli of both $\lambda^*_{i11}\tilde{\lambda}'_{i21}$ and $\lambda'^*_{11i}\tilde{\lambda}'_{12i}$ as well as the RPV weak phase of $\lambda'^*_{11i}\tilde{\lambda}'_{12i}$ are constrained by
 current experimental measurements.
We get $|\lambda^*_{i11}\tilde{\lambda}'_{i21}|\leq1.15\times10^{-2}$ and $|\lambda'^*_{11i}\tilde{\lambda}'_{12i}|\leq7.66$. The bound on slepton exchange coupling $\lambda^*_{i11}\tilde{\lambda}'_{i21}$ is derived for the  first time.
 Squark exchange coupling $\tilde{\lambda}'^*_{11i}\tilde{\lambda}'_{12i}$ gives contribution to $c\to u e^+e^-$ transition and $D^0-\bar{D}^0$ mixing.
There are much stronger bounds from $c\to u e^+e^-$ transition and $D^0-\bar{D}^0$ mixing, which are $|\tilde{\lambda}'^*_{11i}\tilde{\lambda}'_{12i}|\leq4.5\times10^{-2}$  from $c\to u e^+e^-$ transition \cite{wang:2014}
 and $|\tilde{\lambda}'^*_{11i}\tilde{\lambda}'_{12i}|\leq1.0\times10^{-2}$ from  $D^0-\bar{D}^0$ lifetime difference \cite{Petrov:2007gp}. If neglecting the difference between
$\lambda'$ and $\tilde{\lambda}'$, our bound on $\lambda'^*_{11i}\tilde{\lambda}'_{12i}$ from $\bar{c}\to \bar{d} e^+\nu_e$ is more than two orders weaker than one from $c\to u e^+e^-$ transition  or  $D^0-\bar{D}^0$ mixing.

\begin{figure}[t]
\begin{center}
\includegraphics[scale=0.55]{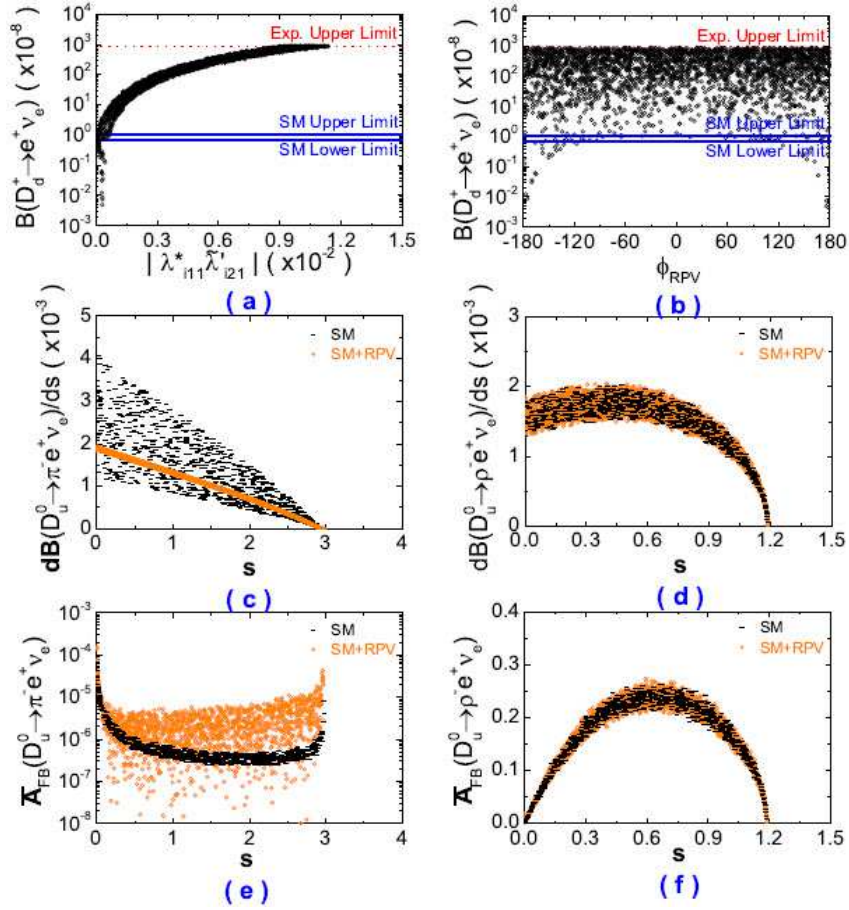}
\end{center}
\vspace{-0.4cm}
 \caption{The constrained effects of RPV coupling $\lambda^*_{i11}\tilde{\lambda}'_{i21}$  due to the slepton exchange in the  exclusive $\bar{c}\to \bar{d}e^+\nu_e$ decays. }
 \label{fig:cdelslp}
\end{figure}

\begin{figure}[t]
\begin{center}
\includegraphics[scale=0.73]{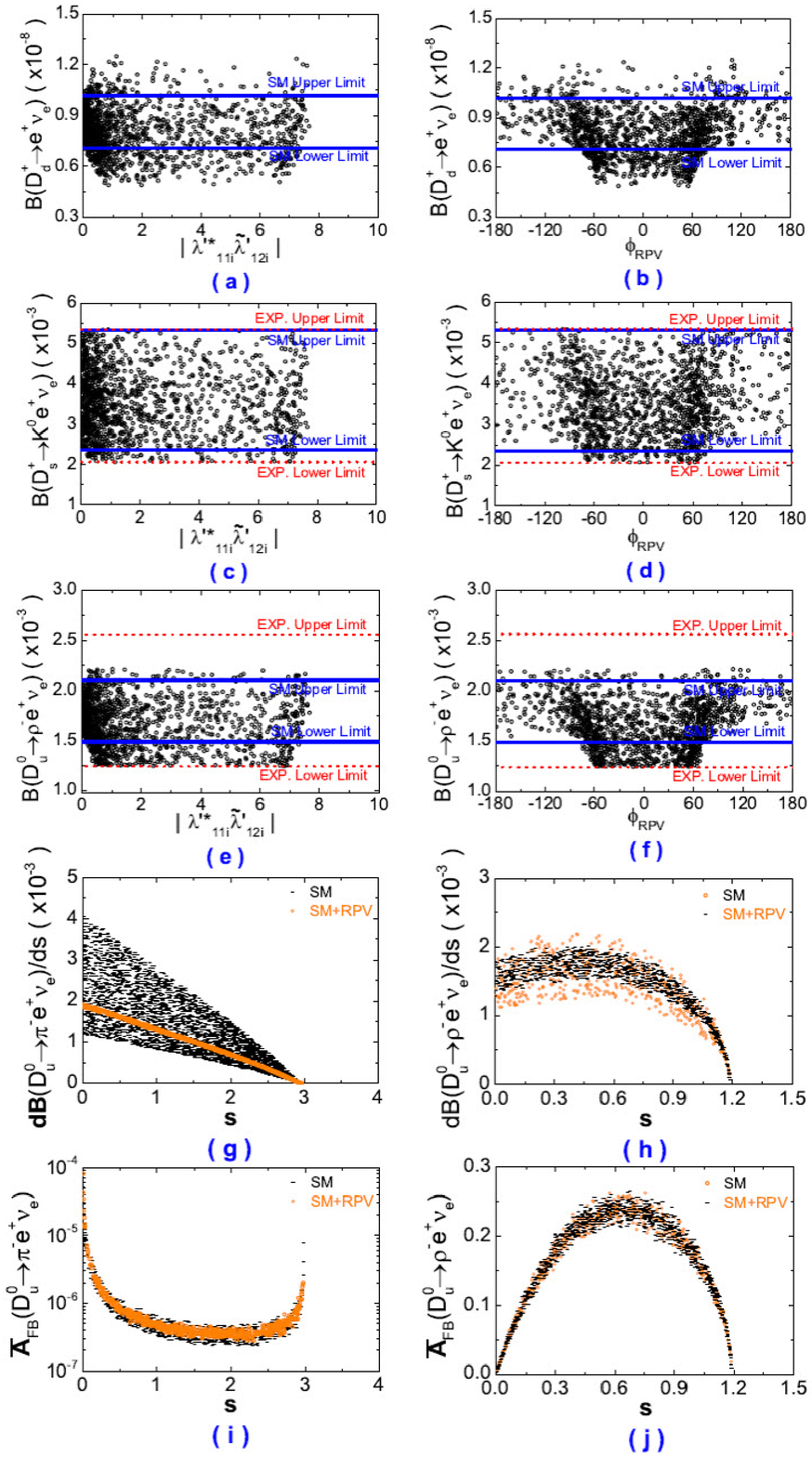}
\end{center}
\vspace{-0.4cm}
 \caption{The constrained effects of RPV coupling $\lambda'^*_{11i}\tilde{\lambda}'_{12i}$ due to the squark exchange in the  exclusive $\bar{c}\to \bar{d}e^+\nu_e$ decays. }
 \label{fig:cdelpslp}
\end{figure}

Now we will analyze the constrained RPV effects in the exclusive $\bar{c} \to \bar{d} e^+ \nu_e$ decays.
Using the constrained parameter spaces shown in Fig. \ref{fig:boundscde}, we can predict the constrained RPV effects on the branching ratios,  the differential branching ratios
and the normalized FB asymmetries of charged leptons. The numerical results for the branching ratios  are listed in the last two columns of Table \ref{tab:cde},
and the constrained RPV effects of $\lambda^*_{i11}\tilde{\lambda}'_{i21}$ and $\lambda'^*_{11i}\tilde{\lambda}'_{12i}$ in the exclusive $ \bar{c} \to \bar{d} e^+ \nu_e$ decays
are displayed in Fig.  \ref{fig:cdelslp} and Fig. \ref{fig:cdelpslp}, respectively.
Comparing the RPV SUSY predictions to the SM ones or experimental bounds
given in Table \ref{tab:cde} as well as shown in Fig.  \ref{fig:cdelslp} and Fig. \ref{fig:cdelpslp}, we give some remarks as follows.

For the slepton exchange  coupling $\lambda^*_{i11}\tilde{\lambda}'_{i21}$, since its contribution to $\mathcal{B}(D^+_d\to e^+\nu_e)$ is increased by  $m_D/m_e$, as shown in  Fig.  \ref{fig:cdelslp} (a-b),  $\mathcal{B}(D^+_d\to e^+\nu_e)$ can be extremely enhanced or reduced  by the constrained $\lambda^*_{i11}\tilde{\lambda}'_{i21}$ coupling, and it is very sensitve to both modulus and weak phase of $\lambda^*_{i11}\tilde{\lambda}'_{i21}$, furthermore,  $|\lambda^*_{i11}\tilde{\lambda}'_{i21}|$ is tightly upper-limited by the experimental measurement of $\mathcal{B}(D^+_d\to e^+\nu_e)$.
 The constrained slepton exchange coupling $\lambda^*_{i11}\tilde{\lambda}'_{i21}$ has no obvious contribution to the six semileptonic decay branching ratios.  From the forth column of Table \ref{tab:cde}, one can find that present accurate experimental measurements of $\mathcal{B}(D^0_u\to \pi^{-}e^+\nu_e)$ and $\mathcal{B}(D^+_d\to \pi^{0}e^+\nu_e)$ give very strong bounds on the semileptonic decay  branching ratio predictions with $\lambda^*_{i11}\tilde{\lambda}'_{i21}$ coupling.
As for the differential branching ratios and  the normalized FB asymmetries  of relevant semileptonic $D$ decays, slepton exchange RPV contributions to $D^0_u\to \pi^{-}e^+\nu_e$, $D^+_d\to \pi^0e^+\nu_e$ and
$D^+_s\to K^0e^+\nu_e$ ($D^0_u\to \rho^{-}e^+\nu_e$, $D^+_d\to \rho^0e^+\nu_e$ and
$D^+_s\to K^{*0}e^+\nu_e$) are very similar to each other. We would take   $D^0_u\to \pi^{-}e^+\nu_e$ and $D^0_u\to \rho^{-}e^+\nu_e$ as
examples (the similar in the subsections of the exclusive $\bar{c}\to \bar{d}\mu^+\nu_\mu, \bar{s}e^+\nu_e,\bar{s}\mu^+\nu_\mu$ decays), which are shown by Fig.  \ref{fig:cdelslp} (c,e) and Fig.  \ref{fig:cdelslp} (d,f), respectively. We can see that present accurate experimental measurements of $\mathcal{B}(D^0_u\to \pi^{-}e^+\nu_e)$ and $\mathcal{B}(D^+_d\to \pi^{0}e^+\nu_e)$ also give very strong bounds on their  differential branching ratios, nevertheless, other differential branching ratios (including $d\mathcal{B}(D^+_s\to K^0e^+\nu_e)/ds$) are not constrained so much by present experimental measurements given in Table \ref{tab:cde}.  The RPV predictions of  the four  differential branching ratios of $D^+_s\to K^0e^+\nu_e$, $D^0_u\to \rho^{-}e^+\nu_e$, $D^+_d\to \rho^0e^+\nu_e$ and
$D^+_s\to K^{*0}e^+\nu_e$ decays can not be distinguished from their SM ones at all $s$ range.
As displayed in Fig.  \ref{fig:cdelslp} (e),  the constrained slepton exchange coupling has quite large effects on  the normalized FB asymmetries  of $D^0_u\to \pi^{-}e^+\nu_e$,  $D^+_d\to \pi^0e^+\nu_e$ and
$D^+_s\to K^0e^+\nu_e$  decays, but these values are very tiny.

The contributions of the squark exchange coupling $\lambda'^*_{11i}\tilde{\lambda}'_{12i}$ are totally different to ones of the slepton exchange coupling $\lambda^*_{i11}\tilde{\lambda}'_{i21}$. Our constrained $\lambda'^*_{11i}\tilde{\lambda}'_{12i}$ has small effects on  $\mathcal{B}(D^+_d\to e^+\nu_e)$, but it has obvious effects in the six semileptonic $D$ decays.
From the last column of Table \ref{tab:cde}, one can see that the experimental measurements of all relevant semileptonic $D$ decays  except $D^+_s\to K^{*0}e^+\nu_e$ give bounds on $\lambda'^*_{11i}\tilde{\lambda}'_{12i}$.
Except $\mathcal{B}(D^0_u\to \pi^{-}e^+\nu_e)$ and $\mathcal{B}(D^+_d\to \pi^{0}e^+\nu_e)$, which are strongly constrained by their experimental measurements, as shown in  Fig.  \ref{fig:cdelpslp} (a-f), all other branching ratios are sensitive to both modulus and weak phase of  $\lambda'^*_{11i}\tilde{\lambda}'_{12i}$.  Note that $\mathcal{B}(D^+_d\to e^+\nu_e)$ including  the constrained  $\lambda'^*_{11i}\tilde{\lambda}'_{12i}$ is much less than its experimental upper limit, therefore we do not show the experimental upper limit in Fig.  \ref{fig:cdelpslp} (a-b).
The constrained squark  exchange contributions to the differential branching ratios and  the normalized FB asymmetries of  $D^0_u\to \pi^{-}e^+\nu_e$, $D^+_d\to \pi^0e^+\nu_e$ and
$D^+_s\to K^0e^+\nu_e$ ($D^0_u\to \rho^{-}e^+\nu_e$, $D^+_d\to \rho^0e^+\nu_e$ and
$D^+_s\to K^{*0}e^+\nu_e$) are very similar to each other. We would also take   $D^0_u\to \pi^{-}e^+\nu_e$ and $D^0_u\to \rho^{-}e^+\nu_e$ as examples (the similar  in the subsections of the exclusive $\bar{c}\to \bar{d}\mu^+\nu_\mu, \bar{s}e^+\nu_e,\bar{s}\mu^+\nu_\mu$ decays), which are displayed in  Fig.  \ref{fig:cdelpslp} (g-j).
Fig.  \ref{fig:cdelpslp} (h-i) show us that our constrained $\lambda'^*_{11i}\tilde{\lambda}'_{12i}$ coupling could enlarge the allowed ranges of $d\mathcal{B}(D^0_u\to \rho^{-}e^+\nu_e)$, but it could shrink the allowed ranges of $\bar{A}_{FB}(D^0_u\to \pi^{-}e^+\nu_e)$.
Noted that, if considering the further constraints from $D^0-\bar{D}^0$ mixing and $c\to u e^+e^-$ transition, i.e., $|\tilde{\lambda}'^*_{11i}\tilde{\lambda}'_{12i}|\leq1.0\times10^{-2}$, the further constrained $\lambda'^*_{11i}\tilde{\lambda}'_{12i}$ coupling  has small effects in the exclusive $\bar{c}\to \bar{d}e^+\nu_e$ decays.

 \subsection{The exclusive $\bar{c}\to \bar{d}\mu^+\nu_\mu$ decays}
 Two RPV coupling products,
$\lambda^*_{i22}\tilde{\lambda}'_{i21}$ due to slepton exchange and $\lambda'^*_{21i}\tilde{\lambda}'_{22i}$ due to squark exchange,  contribute to seven exclusive
$\bar{c}\to \bar{d}\mu^+\nu_\mu$ decay modes,
 $D^+_d\to \mu^+\nu_\mu$, $D^0_u\to\pi^-\mu^+\nu_\mu$,
 $D^+_d\to\pi^0\mu^+\nu_\mu$, $D^+_s\to K^0\mu^+\nu_\mu$,  $D^0_u\to\rho^-\mu^+\nu_\mu$,
 $D^+_d\to\rho^0\mu^+\nu_\mu$ and $D^+_s\to K^{*0}\mu^+\nu_\mu$.
Three relevant branching ratios of  the exclusive $\bar{c}\to \bar{d}\mu^+\nu_\mu$ decays have been  measured by
BESIII \cite{Ablikim:2013uvu}, CLEO-c \cite{Eisenstein:2008aa,Yelton:2009aa},  Belle \cite{Widhalm:2006wz} and BES \cite{Ablikim:2004ry}, and
  their average values from PDG \cite{PDG2014} and  corresponding experimental bound at 90\% CL are given in the second column of Table \ref{tab:cdmu}. The SM prediction values with $1\sigma$ error ranges for
the input parameters  are listed in  the third column of Table \ref{tab:cdmu}.
 \begin{table}[t]
\caption{Branching ratios of the exclusive $\bar{c}\to \bar{d}\mu^+\nu_\mu $ decays (in units of $10^{-3}$) except
  for  $ \mathcal{B}(D^+_d\to \mu^+\nu_\mu)$ (in units of $10^{-4})$. ``a" denotes the experimental data and ``b" denotes the corresponding experimental bound at 90\% CL. }
\begin{center}
\begin{tabular}{lcccc}
\hline\hline
 Observable& Exp. data & SM predictions& SUSY w/$\lambda^*_{i22}\tilde{\lambda}'_{i21}$ & SUSY w/$\lambda'^*_{21i}\tilde{\lambda}'_{22i}$ \\\hline
$\mathcal{B}(D^+_d\to \mu^+\nu_\mu) $&$^{(3.82\pm0.33)^a}_{[3.28,~4.36]^b}$&$[3.02,4.32]$&$[3.38,4.10]$&$[3.38,4.10]$\\\hline
$\mathcal{B}(D^0_u\to\pi^-\mu^+\nu_\mu)$&$^{(2.37\pm0.24)^a}_{[1.98,~2.76]^b}$&$[1.88,6.17]$&$[1.98,2.76]$&$[1.98,2.76]$\\\hline
$\mathcal{B}(D^+_d\to \pi^0\mu^+\nu_\mu)$&$\cdots$&$[2.42,7.92]$&$[2.53,3.58]$&$[2.53,3.58]$\\\hline
$\mathcal{B}(D^+_s\to K^0\mu^+\nu_\mu)$&$\cdots$&$[2.40,5.19]$&$[2.32,5.16]$&$[1.99,6.68]$\\\hline
$\mathcal{B}(D^0_u\to \rho^-\mu^+\nu_\mu)$&$\cdots$&$[1.55,2.15]$&$[1.52,2.17]$&$[1.35,2.37]$\\\hline
$\mathcal{B}(D^+_d\to\rho^0\mu^+\nu_\mu)$&$^{(2.4\pm0.4)^a}_{[1.74,~3.06]^b}$&$[2.01,2.79]$&$[1.97,2.81]$&$[1.74,3.06]$\\\hline
$\mathcal{B}(D^+_s\to K^{*0}\mu^+\nu_\mu)$&$\cdots$&$[1.99,2.79]$&$[2.00,2.80]$&$[1.67,3.45]$\\\hline
\hline
\end{tabular}
\end{center}
\label{tab:cdmu}
\end{table}

\begin{figure}[t]
\begin{center}
\includegraphics[scale=0.65]{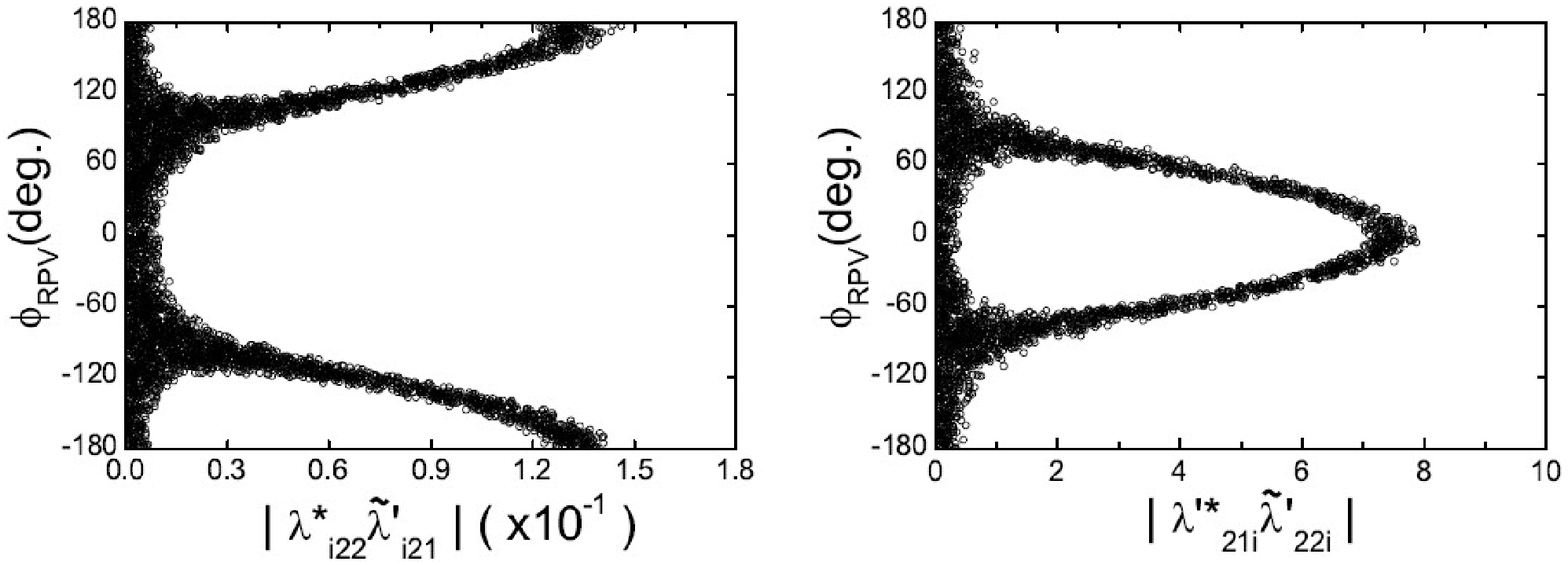}
\end{center}
\vspace{-0.4cm}
 \caption{The allowed RPV parameter spaces from the exclusive $\bar{c}\to \bar{d}\mu^+\nu_\mu$ decays  at 90\% CL with  $500$ GeV sfermion mass.}
 \label{fig:boundscdmu}
\end{figure}

Our bounds for
$\lambda^*_{i22}\tilde{\lambda}'_{i21}$ and $\lambda'^*_{21i}\tilde{\lambda}'_{22i}$  from
the 90\% CL experimental data are demonstrated in
Fig. \ref{fig:boundscdmu},
and  both modulus and weak phase of $\lambda^*_{i22}\tilde{\lambda}'_{i21}$  are strongly constrained by
 the experimental measurements of $\mathcal{B}(D^0_u\to\pi^-\mu^+\nu_\mu)$,  and $\lambda'^*_{21i}\tilde{\lambda}'_{22i}$ is constrained by
 the experimental measurements of $\mathcal{B}(D^0_u\to\pi^-\mu^+\nu_\mu)$ as well as $\mathcal{B}(D^+_d\to\rho^0\mu^+\nu_\mu)$.
 We get $|\lambda^*_{i22}\tilde{\lambda}'_{i21}|\leq1.46\times10^{-1}$ and $|\lambda'^*_{21i}\tilde{\lambda}'_{22i}|\leq7.89$. The bound on slepton exchange coupling $\lambda^*_{i22}\tilde{\lambda}'_{i21}$ is derived for the  first time.
 Squark exchange coupling $\tilde{\lambda}'^*_{21i}\tilde{\lambda}'_{22i}$  could give contribution to $c\to u \mu^+\mu^-$ transition and $D^0-\bar{D}^0$ mixing.
There are much stronger bounds from $c\to u \mu^+\mu^-$ transition and $D^0-\bar{D}^0$ mixing, which are $|\tilde{\lambda}'^*_{21i}\tilde{\lambda}'_{22i}|\leq1.25\times10^{-2}$ from $c\to u \mu^+\mu^-$ transition \cite{wang:2014}
 and $|\tilde{\lambda}'^*_{21i}\tilde{\lambda}'_{22i}|\leq7.25\times10^{-2}$ from  $D^0-\bar{D}^0$ lifetime difference \cite{Petrov:2007gp}. If neglecting the difference between
$\lambda'$ and $\tilde{\lambda}'$, our bound on $\lambda'^*_{21i}\tilde{\lambda}'_{22i}$ from $\bar{c}\to \bar{d} \mu^+\nu_\mu$ is much weaker than   one from $c\to u \mu^+\mu^-$ transition  or  $D^0-\bar{D}^0$ mixing.

\begin{figure}[t]
\begin{center}
\includegraphics[scale=0.55]{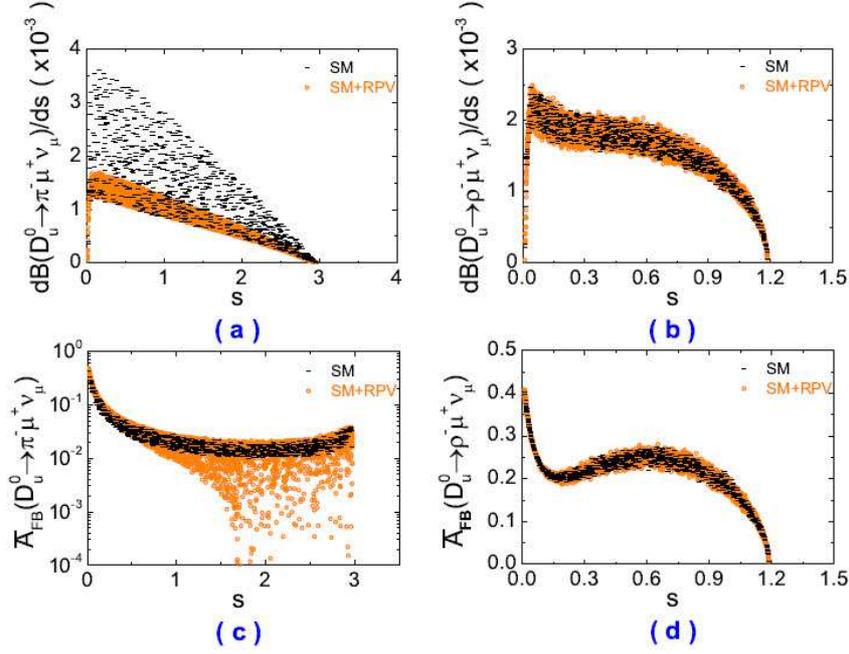}
\end{center}
\vspace{-0.4cm}
 \caption{The constrained effects of RPV coupling $\lambda^*_{i22}\tilde{\lambda}'_{i21}$  due to the slepton exchange in the  exclusive $\bar{c}\to \bar{d}\mu^+\nu_\mu$ decays. }
 \label{fig:cdmulslp}
\end{figure}

\begin{figure}[t]
\begin{center}
\includegraphics[scale=0.70]{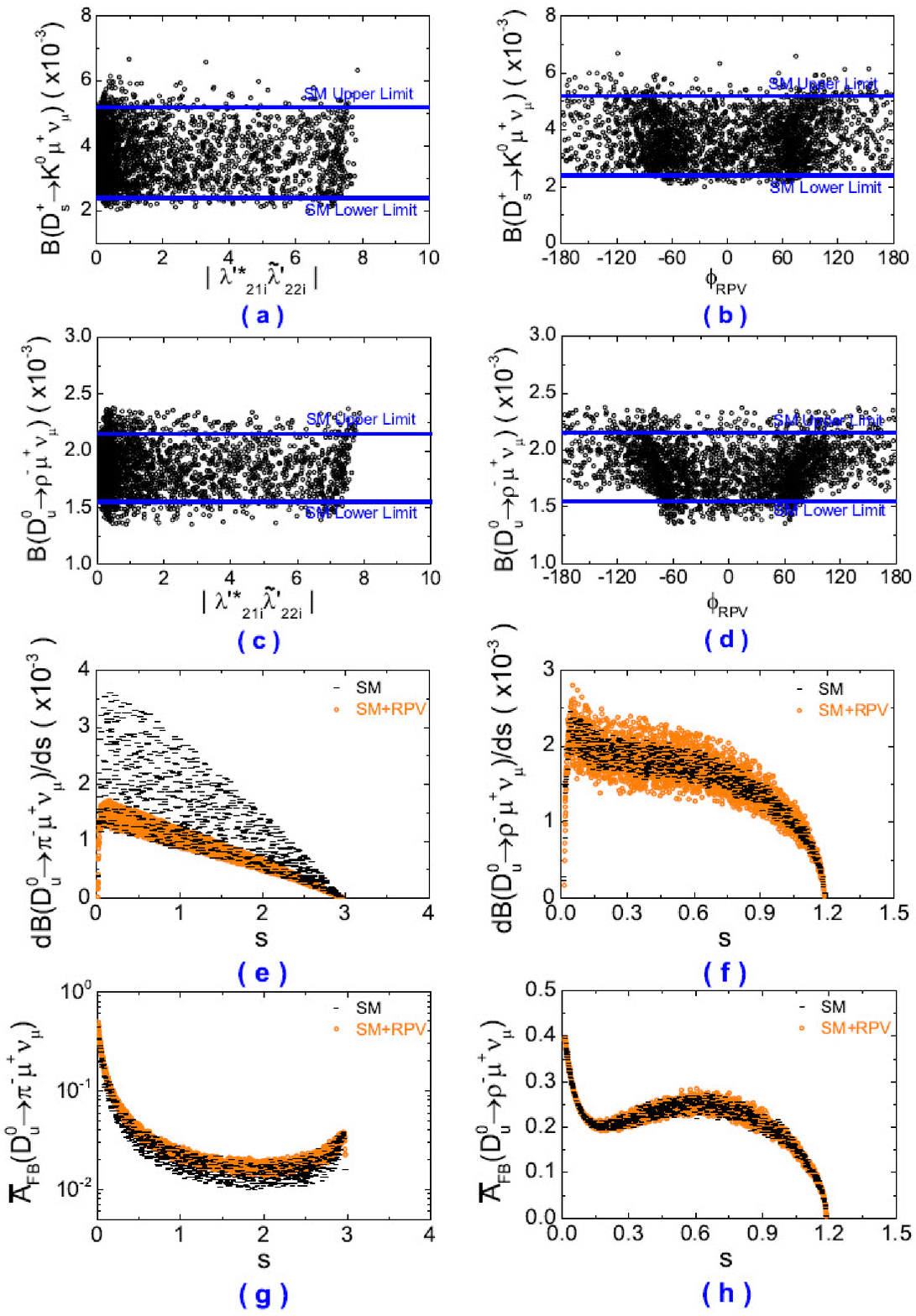}
\end{center}
\vspace{-0.4cm}
 \caption{The constrained effects of RPV coupling $\lambda'^*_{21i}\tilde{\lambda}'_{22i}$ due to the squark exchange in the  exclusive $\bar{c}\to \bar{d}\mu^+\nu_\mu$ decays. }
 \label{fig:cdmulpslp}
\end{figure}

Now we  discuss  the constrained RPV effects in the exclusive $\bar{c} \to \bar{d} \mu^+ \nu_\mu$ decays.
 The numerical results for the branching ratios  are listed in the last two columns of Table \ref{tab:cdmu},
and the constrained RPV effects of $\lambda^*_{i22}\tilde{\lambda}'_{i21}$ and $\lambda'^*_{21i}\tilde{\lambda}'_{22i}$ in the exclusive $\bar{c} \to \bar{d} \mu^+ \nu_\mu$ decays
are shown in Fig.  \ref{fig:cdmulslp} and Fig. \ref{fig:cdmulpslp}, respectively. We have the following remarks for the constrained RPV effects.

For the slepton exchange  coupling $\lambda^*_{i22}\tilde{\lambda}'_{i21}$, since all seven relevant  branching ratios are not sensitive to both modulus and weak phase of the constrained $\lambda^*_{i22}\tilde{\lambda}'_{i21}$ coupling, we do not show them in Fig.  \ref{fig:cdmulslp}.
From the forth column of Table \ref{tab:cdmu} and Fig.  \ref{fig:cdmulslp} (a-b), one can find that present accurate experimental measurements of $\mathcal{B}(D^0_u\to \pi^{-}\mu^+\nu_\mu)$ give very strong bounds on the branching ratios and the differential branching ratios of $D^0_u\to \pi^{-}\mu^+\nu_\mu$ and $D^+_d\to \pi^{0}\mu^+\nu_\mu$ decays, but it does not give obvious bound on other branching ratios and differential branching ratios. As shown in Fig.  \ref{fig:cdmulslp} (c), the constrained $\lambda^*_{i22}\tilde{\lambda}'_{i21}$ coupling could have great effects on $\mathcal{A}_{FB}(D^0_u\to \pi^{-}\mu^+\nu_\mu,D^+_d\to \pi^{0}\mu^+\nu_\mu,D^0+_s\to K^{0}\mu^+\nu_\mu)$  at the middle and high $s$ region, nevertheless, Fig.  \ref{fig:cdmulslp} (d) shows us   the constrained $\lambda^*_{i22}\tilde{\lambda}'_{i21}$ coupling has no obvious effect on $\mathcal{A}_{FB}(D^0_u\to \rho^{-}\mu^+\nu_\mu,D^+_d\to \rho^{0}\mu^+\nu_\mu,D^+_s\to K^*\mu^+\nu_\mu)$.

As for the squark exchange coupling $\lambda'^*_{21i}\tilde{\lambda}'_{22i}$, the examples of the branching ratios are shown in Fig. \ref{fig:cdmulpslp} (a-d),
relevant semileptonic branching ratios expect $\mathcal{B}(D^0_u\to \pi^{-}\mu^+\nu_\mu)$ and $\mathcal{B}(D^+_d\to \pi^{0}\mu^+\nu_\mu)$ are sensitive to both modulus and weak phase of  $\lambda'^*_{21i}\tilde{\lambda}'_{22i}$, and these branching ratios could have minimum at $\phi_{RPV}\in[-70^\circ,70^\circ]$.
From Fig. \ref{fig:cdmulpslp} (e-f), we can see that the constrained $\lambda'^*_{21i}\tilde{\lambda}'_{22i}$ coupling let $\mathcal{B}(D^0_u\to \pi^{-}\mu^+\nu_\mu)$ and $\mathcal{B}(D^+_d\to \pi^{0}\mu^+\nu_\mu)$ close to their SM lower limits, and could enlarge the allowed ranges of other four semileptonic  differential branching ratios.
Fig. \ref{fig:cdmulpslp} (g-h) shows us that the constrained $\lambda'^*_{21i}\tilde{\lambda}'_{22i}$ coupling let $\mathcal{A}_{FB}(D^0_u\to \pi^{-}\mu^+\nu_\mu,D^+_d\to \pi^{0}\mu^+\nu_\mu,D^+_s\to K^{0}\mu^+\nu_\mu)$ close to their SM upper limits, but has no very obvious effect on $\mathcal{A}_{FB}(D^0_u\to \rho^{-}\mu^+\nu_\mu,D^+_d\to \rho^{0}\mu^+\nu_\mu,D^+_s\to K^{*0}\mu^+\nu_\mu)$.
However, if considering the further constraints from $D^0-\bar{D}^0$ mixing and $c\to u \mu^+\mu^-$ transition, the further constrained $\lambda'^*_{21i}\tilde{\lambda}'_{22i}$ coupling  has no obviuos effects in the exclusive $\bar{c}\to \bar{d}\mu^+\nu_\mu$ decays.

 \subsection{The exclusive $\bar{c}\to \bar{s}e^+\nu_e$ decays}
 There are two RPV coupling products,
$\lambda^*_{i11}\tilde{\lambda}'_{i22}$ due to slepton exchange and $\lambda'^*_{12i}\tilde{\lambda}'_{12i}$ due to squark exchange,  contributing to six exclusive
$\bar{c}\to \bar{s}e^+\nu_e$ decay modes,
 $D^+_s\to e^+\nu_e$, $D^0_u\to K^-e^+\nu_e$,
 $D^+_d\to K^0e^+\nu_e$,  $D^0_u\to K^{*-}e^+\nu_e$,
 $D^+_d\to K^{*0}e^+\nu_e$ and $D^+_s\to \phi e^+\nu_e$.
All  relevant semilpetonic branching ratios of  the exclusive $\bar{c}\to \bar{s}e^+\nu_e$ decays have been accurately measured and  the pureleptonic branching ratios of  $D^+_s\to e^+\nu_e$ has been upperlimited by BESIII \cite{Li:2012tr}, CLEO-c \cite{Besson:2009uv,Briere:2010zc,Coan:2005iu,Alexander:2009ux,Ecklund:2009aa},  BABAR \cite{delAmoSanchez:2010jg,Aubert:2008rs}, BES \cite{Ablikim:2006ah,Ablikim:2004ej},  Belle \cite{Widhalm:2006wz} and MARK-III \cite{Adler:1989rw}.
  Their average values from PDG \cite{PDG2014} and  corresponding experimental bounds at 90\% CL are given in the second column of Table \ref{tab:cse}. Moreover, the SM prediction values with $1\sigma$ error ranges for
the input parameters  are listed in  the third column of Table \ref{tab:cse}.
 \begin{table}[b]
\caption{Branching ratios of the exclusive $\bar{c}\to \bar{s}e^+\nu_e $ decays (in units of $10^{-2}$) except
  for  $ \mathcal{B}(D^+_s\to e^+\nu_e)$ (in units of $10^{-7})$. ``a" denotes the experimental data and ``b" denotes the corresponding experimental bounds at 90\% CL. }
\begin{center}
\begin{tabular}{lcccc}
\hline\hline
 Observable& Exp. data & SM predictions& SUSY w/$\lambda^*_{i11}\tilde{\lambda}'_{i22}$ & SUSY w/$\lambda'^*_{12i}\tilde{\lambda}'_{12i}$ \\\hline
$\mathcal{B}(D^+_s\to e^+\nu_e) $&$<1200$&$[1.06,1.44]$&$[0.08,1200]$&$[1.07,1.70]$\\\hline
$\mathcal{B}(D^0_u\to K^-e^+\nu_e)$&$^{(3.55\pm0.05)^a}_{[3.47,3.63]^b}$&$[2.14,4.02]$&$[3.47,3.63]$&$[3.47,3.63]$\\\hline
$\mathcal{B}(D^+_d\to K^0e^+\nu_e)$&$^{(8.83\pm0.22)^a}_{[8.47,9.19]^b}$&$[5.47,10.23]$&$[8.82,9.19]$&$[8.82,9.19]$\\\hline
$\mathcal{B}(D^0_u\to K^{*-}e^+\nu_e)$&$^{(2.16\pm0.16)^a}_{[1.90,2.42]^b}$&$[1.72,2.21]$&$[2.05,2.21]$&$[2.07,2.28]$\\\hline
$\mathcal{B}(D^+_d\to K^{*0}e^+\nu_e)$&$^{(5.52\pm0.15)^a}_{[5.27,5.77]^b}$&$[4.36,5.65]$&$[5.27,5.63]$&$[5.27,5.77]$\\\hline
$\mathcal{B}(D^+_s\to \phi e^+\nu_e)$&$^{(2.49\pm0.14)^a}_{[2.26,2.72]^b}$&$[1.97,2.95]$&$[2.26,2.72]$&$[2.26,2.72]$\\\hline
\hline
\end{tabular}
\end{center}
\label{tab:cse}
\end{table}

\begin{figure}[t]
\begin{center}
\includegraphics[scale=0.65]{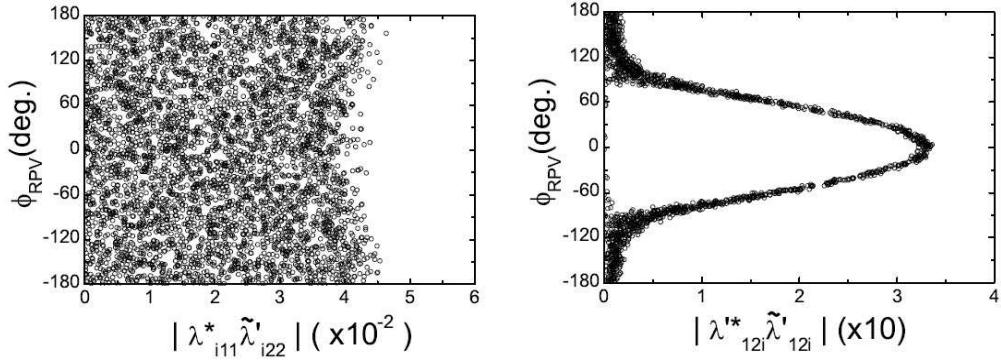}
\end{center}
\vspace{-0.4cm}
 \caption{The allowed RPV parameter spaces from the exclusive $\bar{c}\to\bar{s}e^+\nu_e$ decays  at 90\% CL with  $500$ GeV sfermion mass.}
 \label{fig:boundscse}
\end{figure}

\begin{figure}[t]
\begin{center}
\includegraphics[scale=0.55]{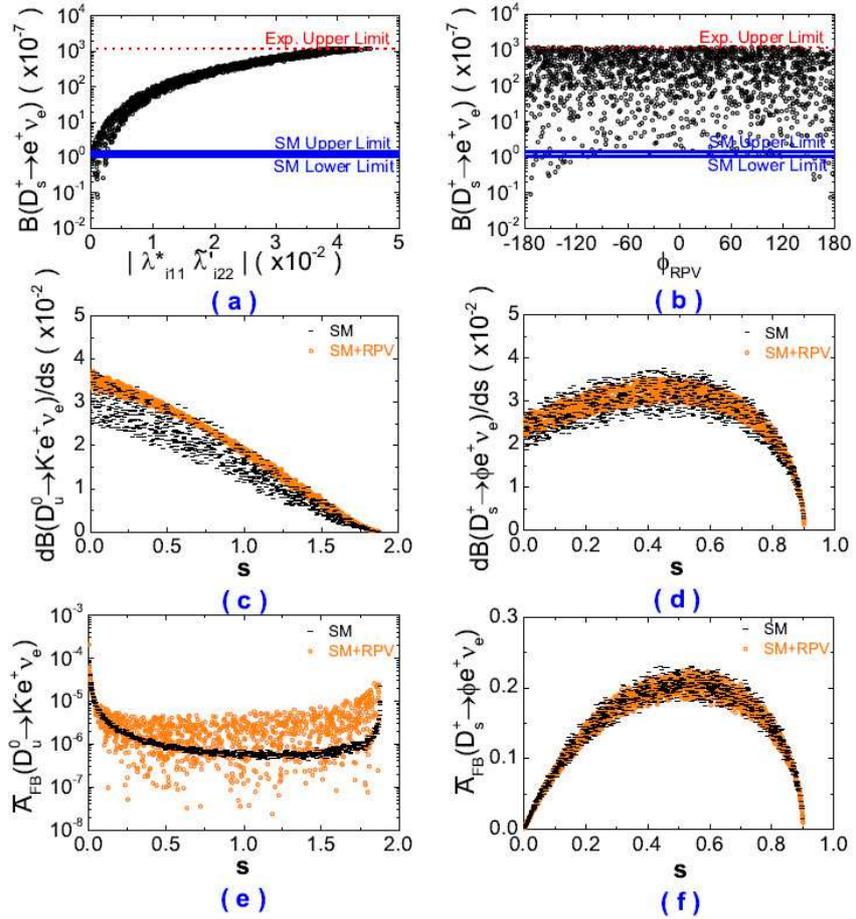}
\end{center}
\vspace{-0.4cm}
 \caption{The constrained effects of RPV coupling $\lambda^*_{i11}\tilde{\lambda}'_{i22}$  due to the slepton exchange in the  exclusive $\bar{c}\to \bar{s}e^+\nu_e$ decays. }
 \label{fig:cselslp}
\end{figure}

\begin{figure}[h]
\begin{center}
\includegraphics[scale=0.55]{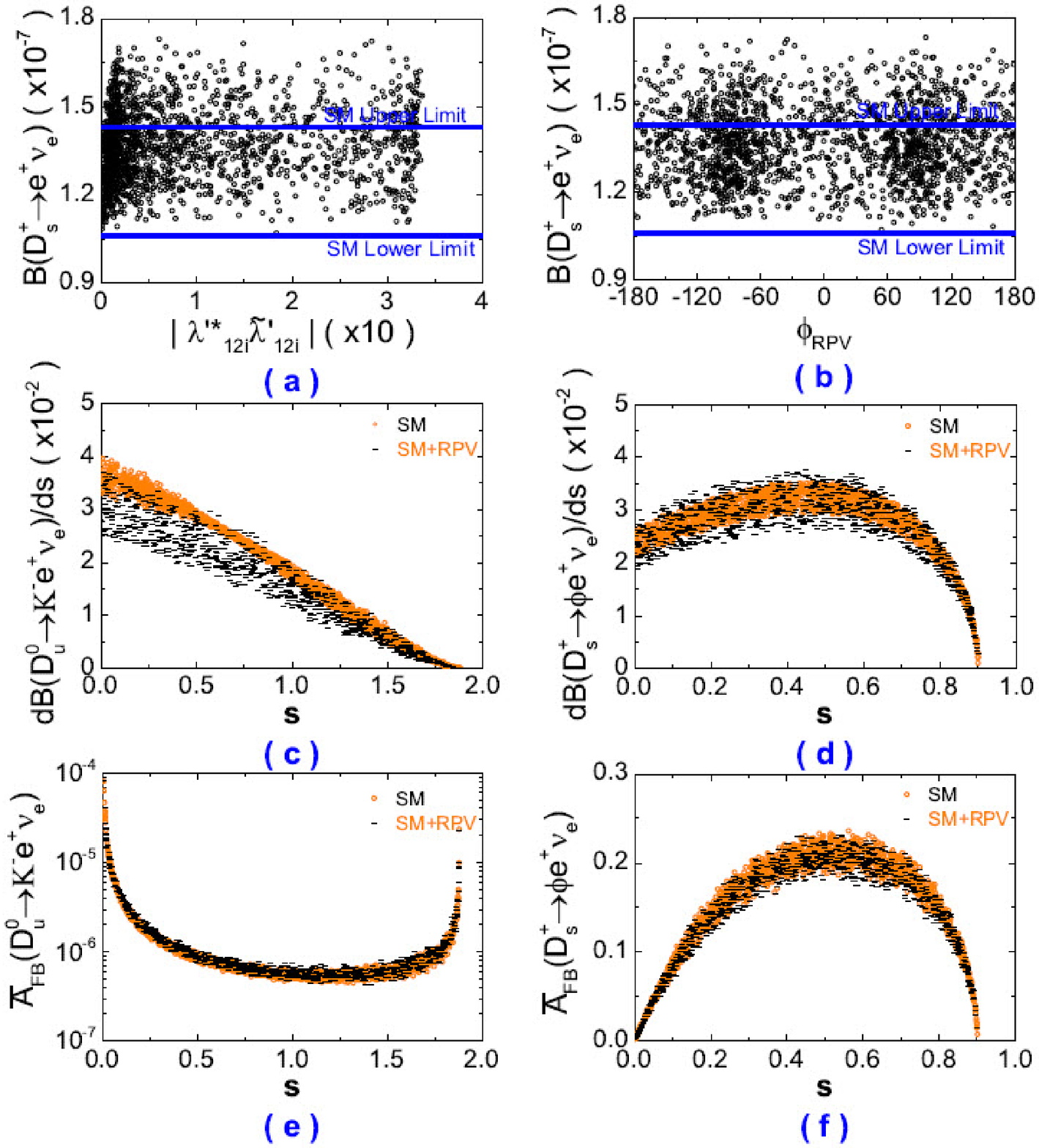}
\end{center}
\vspace{-0.4cm}
 \caption{The constrained effects of RPV coupling $\lambda'^*_{12i}\tilde{\lambda}'_{12i}$ due to the squark exchange in the  exclusive $\bar{c}\to \bar{s}e^+\nu_e$ decays. }
\label{fig:cselpslp}
\end{figure}

Our bounds for
$\lambda^*_{i11}\tilde{\lambda}'_{i22}$ and $\lambda'^*_{12i}\tilde{\lambda}'_{12i}$  from
the 90\% CL experimental data are demonstrated in
Fig. \ref{fig:boundscse}.  We get $|\lambda^*_{i11}\tilde{\lambda}'_{i22}|\leq4.53\times10^{-2}$ and $|\lambda'^*_{12i}\tilde{\lambda}'_{12i}|\leq33.50$. Noted that  both bounds are derived for the  first time.
We also predict the constrained RPV effects in the exclusive $\bar{c}\to\bar{s}e^+\nu_e$ decays. The numerical results for the branching ratios are listed in the last two columns of Table \ref{tab:cse}.  The constrained RPV effects due to the slepton exchange and squark exchange are displayed in  Fig.  \ref{fig:cselslp} and Fig. \ref{fig:cselpslp}, respectively.
One can see that the RPV effects in the exclusive $\bar{c}\to\bar{s}e^+\nu_e$ decays are similar to ones  in  the exclusive $\bar{c}\to \bar{d}e^+\nu_e$ decays.


 \subsection{The exclusive $\bar{c}\to \bar{s}\mu^+\nu_\mu$ decays}
Slepton exchange coupling
$\lambda^*_{i22}\tilde{\lambda}'_{i22}$ and squark exchange coupling $\lambda'^*_{22i}\tilde{\lambda}'_{22i}$   contribute to six exclusive
$\bar{c}\to \bar{s}\mu^+\nu_\mu$ decay modes,
 $D^+_s\to \mu^+\nu_\mu$, $D^0_u\to K^-\mu^+\nu_\mu$,
 $D^+_d\to K^0\mu^+\nu_\mu$,  $D^0_u\to K^{*-}\mu^+\nu_\mu$,
 $D^+_d\to K^{*0}\mu^+\nu_\mu$ and $D^+_s\to \phi \mu^+\nu_\mu$.
All   branching ratios of  the exclusive $c\to s\mu^+\nu_\mu$ decays except $\mathcal{B}(D^+_s\to \phi \mu^+\nu_\mu)$ have been accurately measured by CLEO-c \cite{Briere:2010zc,Alexander:2009ux}, BESII \cite{Ablikim:2006bw}, Belle \cite{Widhalm:2006wz,Widhalm:2007ws}, BABAR \cite{delAmoSanchez:2010jg} and ALEPH \cite{Heister:2002fp}.
 And their average values from PDG \cite{PDG2014} and the SM prediction values are listed in  the second and third columns of Table \ref{tab:csmu}, respectively.
 \begin{table}[ht]
\caption{Branching ratios of the exclusive $\bar{c}\to \bar{s}\mu^+\nu_\mu $ decays (in units of $10^{-2}$) except
  for  $ \mathcal{B}(D^+_s\to \mu^+\nu_\mu)$ (in units of $10^{-3})$. ``a" denotes the experimental data and ``b" denotes the corresponding experimental bounds at 90\% CL. }
\begin{center}
\begin{tabular}{lcccc}
\hline\hline
 Observable& Exp. data & SM predictions& SUSY w/$\lambda^*_{i22}\tilde{\lambda}'_{i22}$ & SUSY w/$\lambda'^*_{22i}\tilde{\lambda}'_{22i}$ \\\hline
$\mathcal{B}(D^+_s\to \mu^+\nu_\mu) $&$^{(5.90\pm0.33)^a}_{[5.36,6.44]^b}$&$[4.52,6.10]$&$[5.36,6.44]$&$[5.36,6.44]$\\\hline
$\mathcal{B}(D^0_u\to K^-\mu^+\nu_\mu)$&$^{(3.31\pm0.13)^a}_{[3.10,3.52]^b}$&$[2.10,5.42]$&$[3.20,3.52]$&$[3.21,3.52]$\\\hline
$\mathcal{B}(D^+_d\to K^0\mu^+\nu_\mu)$&$^{(9.2\pm0.6)^a}_{[8.22,10.18]^b}$&$[5.37,9.94]$&$[8.22,9.03]$&$[8.22,9.03]$\\\hline
$\mathcal{B}(D^0_u\to K^{*-}\mu^+\nu_\mu)$&$^{(1.91\pm0.24)^a}_{[1.52,2.30]^b}$&$[1.79,2.32]$&$[1.96,2.19]$&$[1.96,2.19]$\\\hline
$\mathcal{B}(D^+_d\to K^{*0}\mu^+\nu_\mu)$&$^{(5.28\pm0.15)^a}_{[5.03,5.53]^b}$&$[4.55,5.91]$&$[5.03,5.53]$&$[5.03,5.53]$\\\hline
$\mathcal{B}(D^+_s\to \phi \mu^+\nu_\mu)$&$\cdots$&$[1.99,2.79]$&$[2.07,3.03]$&$[2.04,3.42]$\\\hline
\hline
\end{tabular}
\end{center}
\label{tab:csmu}
\end{table}

\begin{figure}[h]
\begin{center}
\includegraphics[scale=0.55]{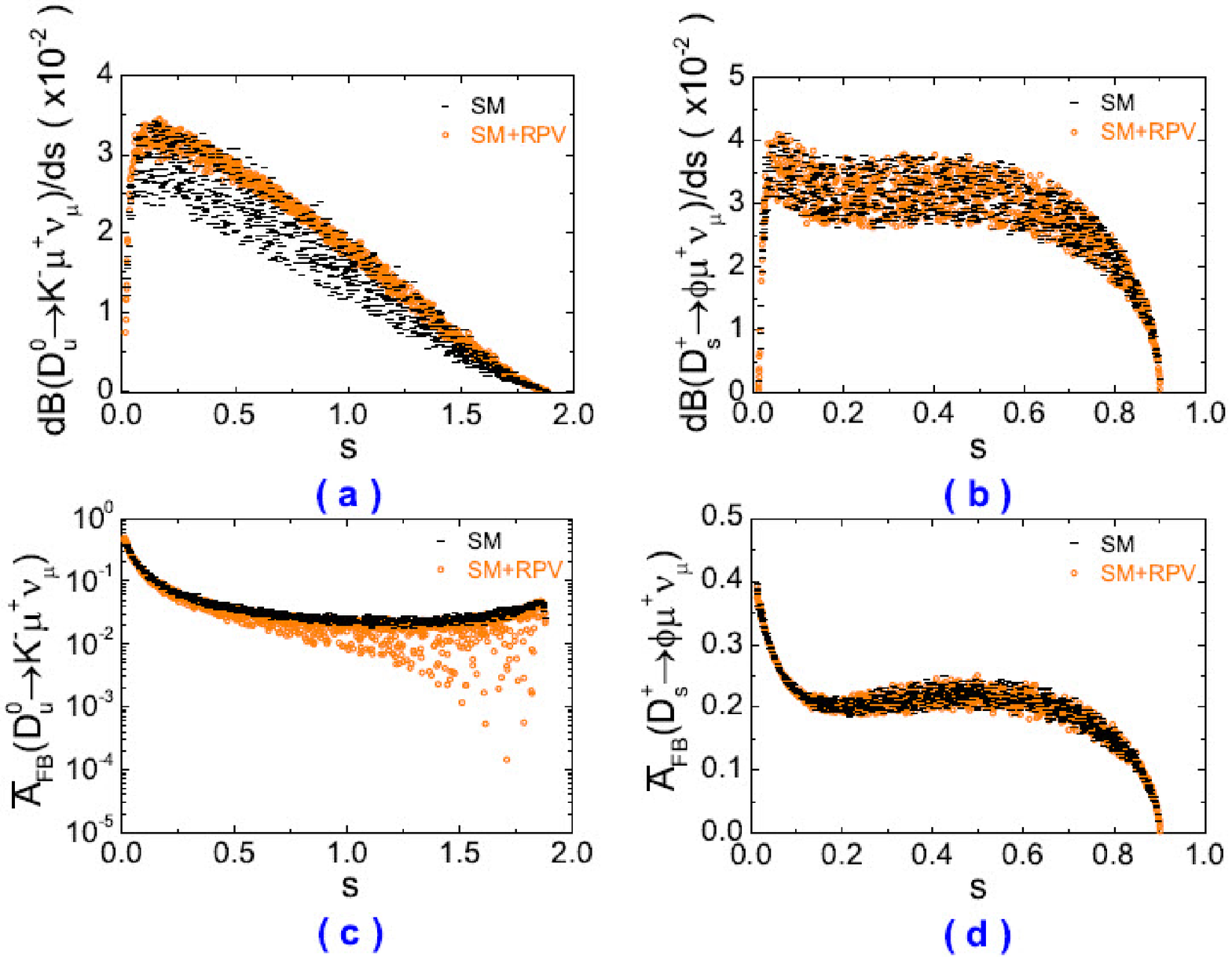}
\end{center}
\vspace{-0.4cm}
 \caption{The constrained effects of RPV coupling $\lambda^*_{i22}\tilde{\lambda}'_{i22}$  due to the slepton exchange in the  exclusive $\bar{c}\to \bar{s}\mu^+\nu_\mu$ decays. }
 \label{fig:csmulslp}
\end{figure}

\begin{figure}[h]
\begin{center}
\includegraphics[scale=0.55]{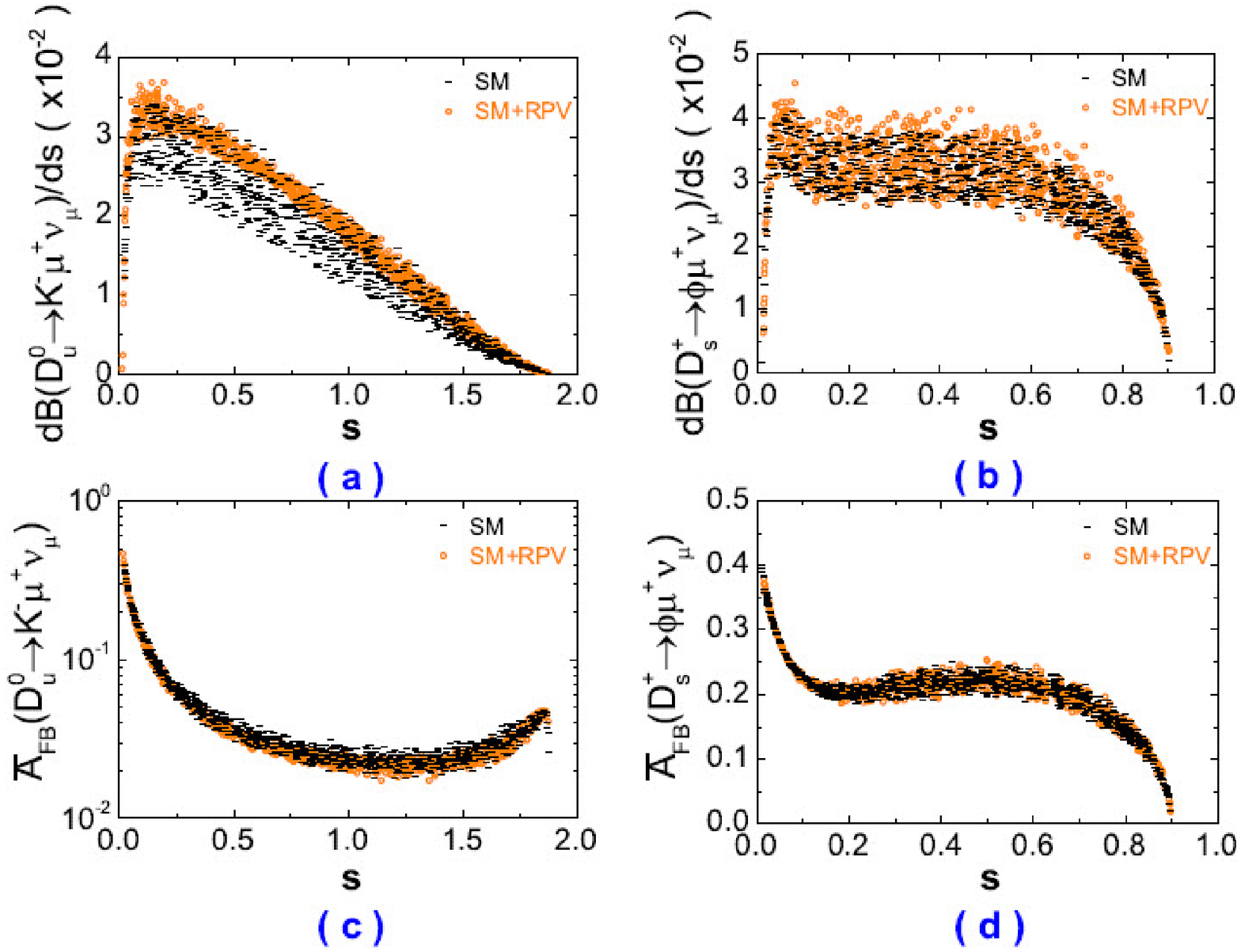}
\end{center}
\vspace{-0.4cm}
 \caption{The constrained effects of RPV coupling $\lambda'^*_{22i}\tilde{\lambda}'_{22i}$ due to the squark exchange in the  exclusive $\bar{c}\to \bar{s}\mu^+\nu_\mu$ decays. }
\label{fig:csmulpslp}
\end{figure}

Our bounds for
$\lambda^*_{i22}\tilde{\lambda}'_{i22}$ and $\lambda'^*_{22i}\tilde{\lambda}'_{22i}$  from
the 90\% CL experimental data are demonstrated in
Fig. \ref{fig:boundscse}.  We get $|\lambda^*_{i22}\tilde{\lambda}'_{i22}|\leq0.60$ and $|\lambda'^*_{12i}\tilde{\lambda}'_{12i}|\leq32.70$, and both bounds are derived for the  first time.
The constrained RPV effects in the exclusive  $\bar{c}\to \bar{s}\mu^+ \nu_\mu$  decays are also explored. The numerical results for the branching ratios are listed in the last two columns of Table \ref{tab:csmu}.  The constrained RPV effects due to the slepton exchange and squark exchange are displayed in  Fig.  \ref{fig:csmulslp} and Fig. \ref{fig:csmulpslp}, respectively.
Noted that the RPV effects in the exclusive $\bar{c}\to\bar{s}\mu^+ \nu_\mu$  decays are similar to ones  in  the exclusive $\bar{c}\to\bar{d}\mu^+\nu_\mu$ decays.
\begin{figure}[h]
\begin{center}
\includegraphics[scale=0.65]{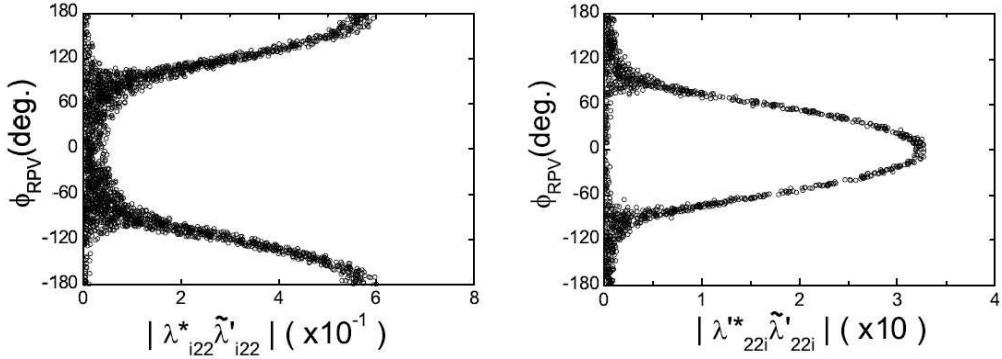}
\end{center}
\vspace{-0.4cm}
 \caption{The allowed RPV parameter spaces from the exclusive $\bar{c}\to \bar{s}\mu^+\nu_\mu$ decays  at 90\% CL with  $500$ GeV sfermion mass.}
 \label{fig:boundscsmu}
\end{figure}

\clearpage
\section{Conclusion}

In this paper, we have studied RPV effects in the 26 semileptonic and leptonic $D$ meson decays, $D_d\to \ell\nu_\ell$, $D_u\to \pi\ell\nu_\ell$, $D_d\to\pi\ell\nu_\ell$, $D_s\to K\ell\nu_\ell$, $D_u\to \rho\ell\nu_\ell$, $D_d\to\rho \ell\nu_\ell$, $D_s\to K^* \ell\nu_\ell$, $D_s\to \ell\nu_\ell$, $D_u\to K\ell\nu_\ell$, $D_d\to K \ell\nu_\ell$, $D_u\to K^{*}\ell\nu_\ell$, $D_d\to K^{*} \ell\nu_\ell$ and $D_s\to \phi \ell\nu_\ell$ with $\ell=e,\mu$.
Considering the theoretical uncertainties and the experimental
errors, we have constrained fairly parameter spaces of RPV coupling constants from the present experimental data, and many obounds are obtained for the first time.
Furthermore, we have predicted the RPV
effects on the branching ratios, the differential branching ratios and  the normalized FB asymmetries of
charged leptons, which have not been measured or
have not been well measured yet.

We have found that the constrained RPV effects due to slepton exchange could be large
on the branching ratios of $D_{d/s}\to e\nu_e$ decays and the normalized FB asymmetries of $D_{u/d}\to \pi/K \ell\nu_\ell$ as well as  $D_s\to K \ell\nu_\ell$ decays.
 The RPV contributions due to squark exchange couplings could enhance the predictions of all semileptonic branching ratios, which are very sensitive to both moduli and weak phases of  the relevent RPV coupling products.
Such correlated signals would
provide strong evidence for RPV interactions. The results in this paper could be useful for probing   the RPV SUSY effects,  and will correlate strongly
with searches for the direct SUSY signals at future experiments.

\section*{Acknowledgments}
The work was supported by the National Natural Science Foundation of
China (No. 11105115
), Joint Funds of the National
Natural Science Foundation of China (No. U1204113),  Program for New
Century Excellent Talents in University (No. NCET-12-0698), and
Natural Research Project of Henan Province (No. 2011A140023).

\begin{appendix}

\section*{Appendix: Input parameters}
\label{SEC.INPUT}
The input parameters except the form factors are collected in Table \ref{table:input parameters}. In our numerical results, we will use the input parameters,which are varied randomly within 1$\sigma$ range.
\begin{table}[ht]
\caption{Default values of the input parameters and the 1$\sigma$ error ranges for the sensitive parameters used in our numerical calculations. }
\vspace{-0.5cm}
\begin{center}
\begin{tabular}{lc}\hline\hline
$m_{K^\pm}=0.493677\pm0.000016~GeV,$~~~$m_{K^0}=0.497614\pm0.000024~GeV,$&\\
$m_{\pi^\pm}=0.13957018\pm0.00000035~GeV,$~~~$m_{\pi^0}=0.1349766\pm0.0000006~GeV,$&\\
$m_{K^{*\pm}}=0.89166\pm0.00026~GeV,$~~~$m_{K^{*0}}=0.89594\pm0.00022~GeV,$&\\
$m_{\rho^\pm}=0.77511\pm0.00034~GeV,$~~~$m_{\rho^0}=0.77526\pm0.00025~GeV,$&\\
$m_{\phi^0}=1.019455\pm0.000020~GeV,~~~\overline{m}_c(\overline{m}_c)=1.275\pm0.0025~GeV,$&\\
$\overline{m}_e(2GeV)=0.0048^{+0.0005}_{-0.0003}~GeV,~~~\overline{m}_s(2GeV)=0.095\pm0.005~GeV,$&\\
$m_W=80.385\pm0.015~GeV,$~~~$m_{c}=1.67\pm0.07~GeV,$&\\
$m_e=0.000510998928\pm0.000000000011~GeV,$~~~$m_{\mu}=0.1056583715\pm0.0000000035~GeV,$&\\
$m_{D_d}=1.86962\pm0.00015~GeV,$~~~$m_{D_s}=1.96850\pm0.00032~GeV,$&\\
$m_{D_u}=1.86486\pm0.00013~GeV.$~~~&\cite{PDG2014}\\\hline
$\tau_{D_u}=0.4101\pm0.0015~ps,~~~\tau_{D_s}=0.500\pm0.007~ps,~~~\tau_{D_d}=1.040\pm0.007~ps.$&\cite{PDG2014}\\\hline
$f_{D_d}=0.201\pm0.017~GeV,~~~f_{D_s}=0.249\pm0.017~GeV.$&\cite{Aubin:2005ar}\\\hline
$|V_{cd}|=0.22520\pm0.0065,~~~|V_{cs}|=0.97344\pm0.00016.$&\cite{PDG2014}\\\hline\hline
\end{tabular}
\end{center}
\label{table:input parameters}
\end{table}

For the form factors involving the $D\rightarrow~P(V)$ transitions, we will use the lightcone $QCD$ sum
rules $(LCSRs)$ results \cite{Wu:2006rd}. For the $s-$dependence of the form factors, they can be parameterized in terms of simple formulae with two or three parameters. To get reasonable behavior of the form factors in the whole kinematically accessible region, we use the following parametrization
\begin{eqnarray}
F(q^2)=\frac{F(0)}{1-a_Fq^2/m^2_M+b_F(q^2/m^2_M)^2},
\end{eqnarray}
where $F(q^2)$ can be any of the form factors $f_+,f_0,A_1,A_2,A_3^{'}$ and $V$. For $D \rightarrow \pi(K)$ decays, we may use the single pole approximations for the form factor $f_+$ in the large $q^2$ region \cite{Wu:2006rd}.
\begin{eqnarray}
f_+(q^2)=\frac{f_{M^*}g_{M^*}M_{\pi}}{2m_{M^*}(1-q^2/m^2_{M^*})},
\end{eqnarray}
with $f_{D^*}g_{D^*}D_{\pi}=2.7\pm0.8GeV,f_{D_s^*}g_{D_s^*}D_{K}=3.1\pm0.6GeV$. With the above considerations,
we obtain the form factors in the whole kinematically accessible region shown in
numerical results are presented in Table \ref{table:form factor}.
\begin{table}[ht]
\caption{Fit for form factors involving the $D \to \pi(K,K^*,\rho)$ and $D_s \to \phi(K,K^*)$ transitions valid for general s \cite{Wu:2006rd}.}
\begin{center}
\begin{tabular}{c|c|c|c|c}\hline\hline
Decay & \multicolumn{2}{c|}{F(0)} & $a_F$ & $b_F$\\
\hline
\multirow{2}*{$D \to \pi$}
&$f_+$&$0.635^{+0.060}_{-0.057}$&$1.01^{-0.12}_{+0.14}$&$0.17^{-0.10}_{+0.13}$\\
\cline{2-5}
&$f_0$&$0.635^{+0.060}_{-0.057}$&$0.64^{-0.01}_{+0.07}$&$-0.20^{-0.04}_{+0.09}$\\
\hline
\multirow{2}*{$D \to K$}
&$f_+$&$0.661^{+0.067}_{-0.066}$&$1.23^{-0.20}_{+0.22}$&$0.69^{-0.15}_{+0.18}$\\
\cline{2-5}
&$f_0$&$0.661^{+0.067}_{-0.066}$&$0.80^{-0.03}_{+0.05}$&$-0.22^{-0.04}_{+0.06}$\\
\hline
\multirow{2}*{$D_s \to K$}
&$f_+$&$0.820^{+0.080}_{-0.071}$&$1.11^{-0.04}_{+0.07}$&$0.49^{-0.05}_{+0.06}$\\
\cline{2-5}
&$f_0$&$0.820^{+0.080}_{-0.071}$&$0.53^{-0.03}_{+0.04}$&$-0.07^{-0.04}_{+0.04}$\\
\hline
\multirow{4}*{$D \to K^*$}
&$A_1$&$0.571^{+0.020}_{-0.022}$&$0.65^{-0.06}_{+0.10}$&$0.66^{-0.18}_{+0.21}$\\
\cline{2-5}
&$A_2$&$0.345^{+0.034}_{-0.037}$&$1.86^{+0.05}_{-0.22}$&$-0.91^{+0.48}_{-0.97}$\\
\cline{2-5}
&$A^{'}_3$&$-0.723^{+0.065}_{-0.077}$&$1.32^{+0.14}_{-0.09}$&$1.28^{+0.22}_{-0.21}$\\
\cline{2-5}
&$V$&$0.791^{+0.024}_{-0.026}$&$1.04^{-0.17}_{+0.25}$&$2.21^{-0.12}_{+0.37}$\\
\hline
\multirow{4}*{$D \to \rho$}
&$A_1$&$0.599^{+0.035}_{-0.030}$&$0.44^{-0.06}_{+0.10}$&$0.58^{-0.04}_{+0.23}$\\
\cline{2-5}
&$A_2$&$0.372^{+0.026}_{-0.031}$&$1.64^{-0.16}_{+0.10}$&$0.56^{-0.28}_{+0.04}$\\
\cline{2-5}
&$A^{'}_3$&$-0.719^{+0.055}_{-0.066}$&$1.05^{+0.15}_{-0.15}$&$1.77^{-0.11}_{+0.20}$\\
\cline{2-5}
&$V$&$0.801^{+0.044}_{-0.036}$&$0.78^{-0.20}_{+0.24}$&$2.61^{+0.29}_{-0.04}$\\
\hline
\multirow{4}*{$D_s \to K^*$}
&$A_1$&$0.589^{+0.040}_{-0.042}$&$0.56^{-0.02}_{+0.02}$&$-0.12^{+0.03}_{-0.02}$\\
\cline{2-5}
&$A_2$&$0.315^{+0.024}_{-0.018}$&$0.15^{+0.22}_{-0.14}$&$0.24^{-0.94}_{+0.83}$\\
\cline{2-5}
&$A^{'}_3$&$-0.675^{+0.027}_{-0.037}$&$0.48^{-0.11}_{+0.13}$&$-0.14^{+0.18}_{-0.17}$\\
\cline{2-5}
&$V$&$0.771^{+0.049}_{-0.049}$&$1.08^{-0.02}_{+0.02}$&$0.13^{+0.03}_{-0.02}$\\
\hline
\multirow{4}*{$D_s \to \phi$}
&$A_1$&$0.569^{+0.046}_{-0.049}$&$0.84^{-0.05}_{+0.06}$&$0.16^{-0.01}_{+0.01}$\\
\cline{2-5}
&$A_2$&$0.304^{+0.021}_{-0.017}$&$0.24^{+0.18}_{-0.05}$&$1.25^{-1.08}_{+1.02}$\\
\cline{2-5}
&$A^{'}_3$&$-0.757^{+0.029}_{-0.039}$&$0.60^{-0.02}_{+0.07}$&$0.60^{+0.31}_{-0.33}$\\
\cline{2-5}
&$V$&$0.778^{+0.057}_{-0.062}$&$1.37^{-0.05}_{+0.04}$&$0.52^{+0.04}_{-0.06}$\\
\hline
\hline
\end{tabular}
\end{center}
\label{table:form factor}
\end{table}

\end{appendix}

\clearpage


\begin{thebibliography}{99}

\bibitem{Ma:2014dha}
  H.~Ma,
  PoS Hadron {\bf 2013}, 076 (2014).

\bibitem{Li:2012tr}
  H.~-B.~Li,
  Nucl.\ Phys.\ Proc.\ Suppl.\  {\bf 233}, 185 (2012)
  [arXiv:1209.3059 [hep-ex]].

\bibitem{Huang:2012qc}
  G.~Huang (BESIII Collaboration),
  arXiv:1209.4813 [hep-ex].


\bibitem{Ablikim:2013uvu}
  M.~Ablikim {\it et al.}  (BESIII Collaboration),
  Phys.\ Rev.\ D {\bf 89}, 051104 (2014)
  [arXiv:1312.0374 [hep-ex]].

\bibitem{Besson:2009uv}
  D.~Besson {\it et al.}  (CLEO Collaboration),
  Phys.\ Rev.\ D {\bf 80}, 032005 (2009)
  [arXiv:0906.2983 [hep-ex]].

\bibitem{Huang:2005iv}
  G.~S.~Huang {\it et al.}  (CLEO Collaboration),
  Phys.\ Rev.\ Lett.\  {\bf 95}, 181801 (2005)
  [hep-ex/0506053].

\bibitem{Coan:2005iu}
  T.~E.~Coan {\it et al.}  (CLEO Collaboration),
  Phys.\ Rev.\ Lett.\  {\bf 95}, 181802 (2005)
  [hep-ex/0506052].

\bibitem{Yelton:2009aa}
  J.~Yelton {\it et al.}  (CLEO Collaboration),
  Phys.\ Rev.\ D {\bf 80}, 052007 (2009)
  [arXiv:0903.0601 [hep-ex]].


\bibitem{Eisenstein:2008aa}
  B.~I.~Eisenstein {\it et al.}  (CLEO Collaboration),
  Phys.\ Rev.\ D {\bf 78}, 052003 (2008)
  [arXiv:0806.2112 [hep-ex]].


\bibitem{Briere:2010zc}
  R.~A.~Briere {\it et al.}  (CLEO Collaboration),
  Phys.\ Rev.\ D {\bf 81}, 112001 (2010)
  [arXiv:1004.1954 [hep-ex]].

 \bibitem{Alexander:2009ux}
  J.~P.~Alexander {\it et al.}  (CLEO Collaboration),
  Phys.\ Rev.\ D {\bf 79}, 052001 (2009)
  [arXiv:0901.1216 [hep-ex]].


 \bibitem{Ecklund:2009aa}
  K.~M.~Ecklund {\it et al.}  (CLEO Collaboration),
  Phys.\ Rev.\ D {\bf 80}, 052009 (2009)
  [arXiv:0907.3201 [hep-ex]].


\bibitem{Widhalm:2006wz}
  L.~Widhalm {\it et al.}  (Belle Collaboration),
  Phys.\ Rev.\ Lett.\   {\bf 97}, 061804 (2006)
  [hep-ex/0604049].

\bibitem{Widhalm:2007ws}
  L.~Widhalm {\it et al.}  (Belle Collaboration),
  Phys.\ Rev.\ Lett.\  {\bf 100}, 241801 (2008)
  [arXiv:0709.1340 [hep-ex]].


 \bibitem{delAmoSanchez:2010jg}
  P.~del Amo Sanchez {\it et al.}  (BaBar Collaboration),
  Phys.\ Rev.\ D {\bf 82}, 091103 (2010)
  [arXiv:1008.4080 [hep-ex]].


 \bibitem{Aubert:2008rs}
  B.~Aubert {\it et al.}  (BaBar Collaboration),
  Phys.\ Rev.\ D {\bf 78}, 051101 (2008)
  [arXiv:0807.1599 [hep-ex]].




\bibitem{Ablikim:2006bw}
  M.~Ablikim {\it et al.}  (BES Collaboration),
  Phys.\ Lett.\ B {\bf 644}, 20 (2007)
  [hep-ex/0610020].


\bibitem{Ablikim:2004ry}
  M.~Ablikim {\it et al.}  (BES Collaboration),
  Phys.\ Lett.\ B {\bf 610}, 183 (2005)
  [hep-ex/0410050].


\bibitem{Ablikim:2004ku}
  M.~Ablikim {\it et al.}  (BES Collaboration),
  Phys.\ Lett.\ B {\bf 608}, 24 (2005)
  [hep-ex/0410030].

 \bibitem{Ablikim:2006ah}
  M.~Ablikim {\it et al.}  (BES Collaboration),
  Eur.\ Phys.\ J.\ C {\bf 47}, 31 (2006)
  [hep-ex/0605103].


 \bibitem{Ablikim:2004ej}
  M.~Ablikim {\it et al.}  (BES Collaboration),
  Phys.\ Lett.\ B {\bf 597}, 39 (2004)
  [hep-ex/0406028].


\bibitem{Heister:2002fp}
  A.~Heister {\it et al.}  (ALEPH Collaboration),
  Phys.\ Lett.\ B {\bf 528}, 1 (2002)
  [hep-ex/0201024].

\bibitem{Adler:1989rw}
  J.~Adler {\it et al.}  (MARK-III Collaboration),
  Phys.\ Rev.\ Lett.\  {\bf 62}, 1821 (1989).

\bibitem{Barranco:2014bva}
  J.~Barranco, D.~Delepine, V.~Gonzalez Macias and L.~Lopez-Lozano,
  arXiv:1404.0454 [hep-ph].

\bibitem{Barranco:2013tba}
  J.~Barranco, D.~Delepine, V.~Gonzalez Macias and L.~Lopez-Lozano,
  Phys.\ Lett.\ B {\bf 731}, 36 (2014)
  [arXiv:1303.3896 [hep-ph]].

\bibitem{Akeroyd:2009tn}
  A.~G.~Akeroyd and F.~Mahmoudi,
  JHEP {\bf 0904}, 121 (2009)
  [arXiv:0902.2393 [hep-ph]].

\bibitem{Dobrescu:2008er}
  B.~A.~Dobrescu and A.~S.~Kronfeld,
  Phys.\ Rev.\ Lett.\  {\bf 100}, 241802 (2008)
  [arXiv:0803.0512 [hep-ph]].

\bibitem{Akeroyd:2007eh}
  A.~G.~Akeroyd and C.~H.~Chen,
  Phys.\ Rev.\ D {\bf 75}, 075004 (2007)
  [hep-ph/0701078].

\bibitem{Fajfer:2006uy}
  S.~Fajfer and J.~F.~Kamenik,
  Phys.\ Rev.\ D {\bf 73}, 057503 (2006)
  [hep-ph/0601028].

\bibitem{Fajfer:2005ug}
  S.~Fajfer and J.~F.~Kamenik,
  Phys.\ Rev.\ D {\bf 72}, 034029 (2005)
  [hep-ph/0506051].


\bibitem{Fajfer:2004mv}
  S.~Fajfer and J.~F.~Kamenik,
  Phys.\ Rev.\ D {\bf 71}, 014020 (2005)
  [hep-ph/0412140].



\bibitem{Akeroyd:2003jb}
  A.~G.~Akeroyd,
  Prog.\ Theor.\ Phys.\  {\bf 111}, 295 (2004)
  [hep-ph/0308260].

\bibitem{Akeroyd:2002pi}
  A.~G.~Akeroyd and S.~Recksiegel,
  Phys.\ Lett.\ B {\bf 554}, 38 (2003)
  [hep-ph/0210376].






\bibitem{Weinberg:1981wj}
  S.~Weinberg,
  Phys.\ Rev.\ D {\bf 26}, 287 (1982).



\bibitem{Chemtob:2004xr}
  M.~Chemtob,
  Prog.\ Part.\ Nucl.\ Phys.\  {\bf 54}, 71 (2005)
  [hep-ph/0406029].

\bibitem{Petrov:2007gp}
  A.~A.~Petrov and G.~K.~Yeghiyan,
  Phys.\ Rev.\ D {\bf 77}, 034018 (2008)
  [arXiv:0710.4939 [hep-ph]].


\bibitem{Golowich:2006gq}
  E.~Golowich, S.~Pakvasa and A.~A.~Petrov,
  Phys.\ Rev.\ Lett.\  {\bf 98}, 181801 (2007)
  [hep-ph/0610039].


\bibitem{Chen:2007dg}
  S.~-L.~Chen, X.~-G.~He, A.~Hovhannisyan and H.~-C.~Tsai,
  JHEP {\bf 0709}, 044 (2007)
  [arXiv:0706.1100 [hep-ph]].



\bibitem{Nandi:2006qe}
  S.~Nandi and J.~P.~Saha,
  Phys.\ Rev.\ D {\bf 74}, 095007 (2006)
  [hep-ph/0608341].

\bibitem{Bhattacharyya:1998be}
  G.~Bhattacharyya and A.~Raychaudhuri,
  Phys.\ Rev.\ D {\bf 57}, 3837 (1998)
  [hep-ph/9712245].

\bibitem{Kundu:2004cv}
  A.~Kundu and J.~P.~Saha,
  Phys.\ Rev.\ D {\bf 70}, 096002 (2004)
  [hep-ph/0403154].






\bibitem{Grinstein:1992qt}
  B.~Grinstein {\it et al.},
  Nucl.\ Phys.\ B {\bf 380}, 369 (1992)
  [hep-ph/9204207].


\bibitem{Wu:2006rd}
  Y.~-L.~Wu, M.~Zhong and Y.~-B.~Zuo,
  Int.\ J.\ Mod.\ Phys.\ A {\bf 21}, 6125 (2006)
  [hep-ph/0604007].


\bibitem{PDG2014}
  J.~Beringer {\it et al.}  (Particle Data Group Collaboration),
  Phys.\ Rev.\ D {\bf 86}, 010001 (2012) and 2013 partial update for the 2014 edition.

\bibitem{wang:2014}
   Ru-Min Wang {\it et al.}, in preparation.

\bibitem{Aubin:2005ar}
  C.~Aubin  {\it et al.},
  Phys.\ Rev.\ Lett.\  {\bf 95}, 122002 (2005)
  [hep-lat/0506030].



\end{thebibliography}
\end{document}